\documentclass[preprint]{ptephy_v1}

\preprintnumber{XXXX-XXXX} 
\usepackage{hyperref}
\usepackage{lineno}

  \usepackage{amsfonts}
\usepackage{rotating} 
\usepackage{epsfig}
\usepackage{multirow}
\usepackage{bm}
\usepackage{amsmath}
\usepackage{graphicx}
\usepackage{hyperref,fancyref}
\usepackage{cleveref}
\usepackage{caption}
\usepackage[toc,page]{appendix}
\usepackage{nameref}
\usepackage{ragged2e}
\usepackage{caption}
\newcommand{\diracslash}[1]{#1\llap{/\kern2pt}}

\newcommand{\be}{\begin{equation}}
	\newcommand{\ee}{\end{equation}}
\newcommand{\bea}{\begin{eqnarray}}
	\newcommand{\eea}{\end{eqnarray}}
\newcommand{\ba}[1]{\begin{array}{#1}}
	\newcommand{\ea}{\end{array}}

\newcommand{\bt}{\begin{tabular}}
	\newcommand{\et}{\end{tabular}}

\newcommand{\beas}{\begin{eqnarray*}}
	\newcommand{\eeas}{\end{eqnarray*}}

\begin{document}

\title{Magnetic moments of $\frac{1}{2}^-$ baryon resonances in strange hadronic matter at high densities}

\author{Abhinaba Upadhyay}

\author{Arvind Kumar}

\author{Harleen Dahiya}

\author{Suneel Dutt \thanks{These authors contributed equally to this work}}
\affil{Department of Physics, Dr. B R Ambedkar National Institute of Technology Jalandhar, 
	Jalandhar -- 144008, Punjab, India
\email{dutts@nitj.ac.in}}

\begin{abstract}%
This work primarily focusses on determining the magnetic moments of $\frac{1}{2}^-$ baryon resonances in the presence of hot and dense hadronic matter. In the chiral $SU(3)$ quark mean field model approach, we have essentially accounted for the effects on in-medium scalar meson fields to investigate the impact of high densities on the in-medium baryon masses and their constituent quarks. In light of chiral constituent quark model $\chi$CQM, we have calculated the magnetic moments of $\frac{1}{2}^-$ baryon resonances and scrutinized the effects due to its internal constituents: the valence quarks, sea quarks and the orbital moment of sea quarks. Furthermore, we have investigated the effective baryonic magnetic moments for the finite magnitudes of isospin asymmetry and strangeness fraction.
\end{abstract}

\subjectindex{xxxx, xxx}

\maketitle

	\section{\label{intro}Introduction}
The theory of quantum chromodynamics (QCD), is based on the consideration of quarks and gluons  as the fundamental degrees of freedom which form bound states known as hadrons (baryons($qqq$) and mesons($q \bar{q}$)). These fundamental degrees of freedom undergo dynamical changes due to strong interactions in dense nuclear medium, which in turn cause in-medium modifications in the properties of hadrons. Confinement makes it challenging to understand QCD at lower energy scales. Studying the internal structure and the properties of light hadrons allow us to unravel the physics of strong interactions at lower energy regimes. Apart from heavy-ion collision experimental studies on chiral symmetry restoration at high temperature \cite{thirtyseven}, and the de-confinement of hadrons to quark-gluon plasma (QGP) \cite{thirtyeight}, it has become a primary interest for experiments such as PANDA at FAIR \cite{thirtythree,thirtyfour},  BESIII at BEPCII \cite{thirtysix,thirtytwo}, HADES at GSI \cite{thirtyfive} etc., to 
explore the electromagnetic properties of hadrons in dense matter. Electromagnetic form factors, being key observable to investigate the electric and magnetic moment distributions, are of theoretical interest to explore other primary hadronic properties like magnetic moment and charged radii to probe deep inside of hadrons. 

Magnetic moment of hadrons, in free space, play a very important role in investigating their internal structure at a sub-nuclear level \cite{sixtyfour}. 
In theory, there has been extensive studies on the magnetic moments of octet and decuplet baryons in free space across a range of different models \cite{one,two,three,four,five,six,seven,eight,nine}. Isgur framed his model with one-gluon exchange, in order to account for internal structure of hadrons \cite{isgur1,isgur2,isgur3,isgur4}. The constituent quark model (CQM) propounded that the baryonic magnetic moments are given as the summation of the individual magnetic moments of its constituent quarks \cite{ten,eleven}. It was Glozman who incorporated the Nambu Goldstone boson exchange (NGBE) interaction in the low energy regime and the concept of one-gluon exchange was substituted by vector meson exchange interactions, which in fact led to better fits for baryon spectra \cite{glozman}. It was not until the development of chiral constituent quark model ($\chi \mathrm{CQM}$) that the pseudo-scalar and scalar meson exchanges were considered which was later extended to the inclusion of vector meson exchange \cite{dai}. Hence,  $\chi \mathrm{CQM}$ was formed by extending the quark model and further incorporating the chiral symmetry and corresponding spontaneous symmetry breaking into it \cite{twelve}. The $\chi \mathrm{CQM}$ model provides satisfactory explanation to the \textit{“Proton spin puzzle”}, which was an outcome of the final data taking at EMC, and other hadron properties \cite{ten,thirteen,fourteen,fifteen,sixteen,seventeen,eighteen,nineteen,twenty,twentyone,twentytwo,twentythree,twentyfour,twentyfive,twentysix,twentyseven}. In previous studies, this model has been used for calculating the baryonic magnetic moments, by taking into account the sea quark polarizations and their orbital angular momentum contribution through the Cheng-Li mechanism \cite{twentyeight,twentynine,thirty}. In the $\chi$CQM formalism, the magnetic moments of low-lying ${\frac{1}{2}}^-$ $N^*$ resonances \cite{thirtyone}, ${\frac{1}{2}}^-$ octet baryon resonances \cite{thirtyone} and low-lying ${\frac{1}{2}}^-$ $\Lambda$ resonances \cite{A_Torres} have been calculated in free space, in earlier studies. Apart from $\chi$CQM, the magnetic moments of $N^*$ resonances have been computed by three quark Hamiltonian approach and the results were not consistent with lattice QCD values with very large variations \cite{hamiltonian_approach}. As an extension to chiral perturbation theory, the Chiral unitary model was also used in ref. \cite{cum_hyodo} to calculate the magnetic moments of $N^* (1535)$ resonances.\par
As mentioned earlier, the study of magnetic moments of baryons and the medium dependent effects on the electromagnetic properties is an area of active research and stands relevant to some of the experiments. For studying the properties of nuclear matter, various theoretical models like the MIT bag model \cite{thirtynine}, linear sigma model \cite{Petro}, non-linear sigma model \cite{fortyfive}, Zimanyi and Moszkowski model \cite{seventyone}, cloudy bag model \cite{fortythree}, Nambu-Jona-Lasinio (NJL) model \cite{fortytwo}, light-front quark model \cite{Schlumpf1993}, parity doublet model \cite{paritydoublet1,paritydoublet2} etc., have been developed. The cloudy bag model attained increased accuracy over MIT bag model in explaining QCD effects by incorporating the presence of pion clouds at the core of a baryon. With further development, models like the NJL model successfully explained low energy baryonic dynamics by considering chiral quarks as constituents of hadrons, through the inclusion of the interaction between the constituent quarks \cite{njlsuccess}. Few years later, the quark meson coupling (QMC) model was constituted and it facilitated a deeper understanding of the in-medium baryon properties \cite{fortyfour}. The QMC model, formulated in lines of cloudy bag and NJL model, has previously been deployed to study the impact of medium on the magnetic moments of octet baryons and has produced results which are in good agreement with experimental data \cite{seventyfour}. The chiral $SU(3)$ quark mean field model (CQMF) \cite{fortyfour,fortyfive,fortysix,fortyseven} is an extended version of the quark mean field model \cite{fortyeight}, which is based on the QMC model approach. In the CQMF model, the constituent quarks are confined into baryons by a confining potential based on mean field approximation using classical expectation values instead of quantum field operators \cite{fortynine}. In the context of medium modifications of electromagnetic properties of hadrons, the studies in ref. \cite{kaon} show the in-medium effects on $K^+$ meson properties in light-front quark meson coupling model (LF-QMC) in symmetric nuclear matter. With the rise in medium density, the kaon electromagnetic form factors decrease faster than in vacuum, while the kaon electromagnetic charge radii grow in size. By studying pion and kaon meson structures in NJL model the authors in ref. \cite{njlmeson} have obtained results consistent with that in ref. \cite{kaon}. For octet baryons in light of the QMC model \cite{effoctet}, it has been shown that the in-medium electric form factors are dampened and the magnetic form factors are enhanced, and this finding is consistent with the in-medium variations of the bound proton electromagnetic form factors observed at Jefferson Lab \cite{jlab}. There has been studies on the effective baryonic magnetic moments aimed to carry out calculation for the low-lying charm and the low-lying bottom baryons, to examine the roles of the heavy and light quarks in symmetric nuclear matter \cite{tsushima1,tsushima2}. While a comparative study on octet baryons with modified quark meson coupling model (MQMC) \cite{mqmc}, suggests that the observed in-medium baryonic magnetic moments incur drastic variations compared to that in QMC framework. In CQMF formalism, the medium modifications of  magnetic moments of octet  \cite{fiftythree}, decuplet \cite{fiftyfour} and charmed \cite{Dutt:2024lui} baryons  have been  studied in hot and dense hadronic matter. 
The CQMF model has recently been used to study valence quark distribution and transverse momentum dependent distributions of valence quarks of pions \cite{Puhan:2024xdq,Kaur:2024wze}.

In the current work we calculate the magnetic moments of the low-lying $N^*$ resonances and $\frac{1}{2}^-$ octet baryon resonances in hot and dense hadronic matter. It is exciting to study these resonances, as their detection connects hadron models to QCD properties in low and intermediate energy regimes. Photoproduction has been a key experimental tool involved in detecting the low-lying states of baryon resonances as shown in refs. \cite{fifty,fiftyone,fiftytwo}. The large decay widths of these excited baryon resonances, makes it challenging to detect them in experiments.
While Ref. \cite{thirtyone} emphasizes on magnetic moment calculations of negative parity octet baryons in free space, our study extends beyond free space in hot and dense hadronic medium. 
As we shall see later, to calculate the magnetic moments of constituent quarks
the impact of relativistic quark confinement is taken into account
in the calculations of present work
\cite{fiftyfive}.
However, in Ref. \cite{thirtyone} 
such confinement effects have not been
considered and this  causes  a difference in the values of vacuum baryonic magnetic moments of this study as compared to our current calculations.
Presence of strangeness fraction ($f_s$) is a principal aspect to understand the departure of hadronic properties of baryons in strange hadronic matter comprising of hyperons,  from that in a non-strange medium. The in-medium variations of magnetic moments of octet and decuplet baryons in strange medium has shown stark deviation, as studied in refs. \cite{fiftythree,fiftyfour,sixtytwo}, for baryons of higher strangeness quantum number. The effects of increasing isospin asymmetry has also been considered for the analysis of density dependent in-medium modifications of baryons.
This manuscript has been organized as follows: \Cref{Model} comprises of two subsections describing  the theoretical models used in the current work. In \cref{sec:chiral_mean_field} the CQMF model used to compute the in-medium quark masses and effective baryon masses, at finite temperature and density of the medium is presented. In \cref{sec:magnetic} we have calculated the contributions of valence quarks, sea quarks and orbital angular momentum of sea quarks towards the total magnetic moments of baryons, in light of $\chi \mathrm{CQM}$ model. \Cref{sec:results} is dedicated to a detailed discussion on the results obtained from numerical calculations. 
We devote \cref{sec:summary} to present the summary of the current work.

\section{Theoretical Framework}\label{Model}
Central to this work, we have used $\chi \mathrm{CQM}$ to calculate the properties of the baryon resonances as in Ref. \cite{thirtyone} and have visualized them as a system of three valence quarks. We have calculated the in-medium magnetic moments of the ${\frac{1}{2}}^-$ baryon resonances at finite temperature and density of hadronic medium through the medium modification of the corresponding constituent quark masses and the baryon masses. Within the CQMF framework, the effective masses of quarks ($m^*_q$) are calculated through the medium modified values of the scalar fields $\sigma$ and $\zeta$ and scalar iso-vector field $\delta$. The confined quarks inside the baryons, participate in the exchange interaction with  these scalar meson fields as well as
vector meson fields $\omega$, $\rho$ and $\phi$. 
\subsection{Chiral Quark Mean Field Model } \label{sec:chiral_mean_field}
In this section, the CQMF model is discussed briefly which serves as the basic framework
to calculate the in-medium masses of constituent quark and baryon. As discussed earlier, in CQMF model the quarks and mesons serve as fundamental degrees of freedom, and the confinement of quarks by an effective confining potential results in the formation of baryons. The constituent quarks and the mesons obtain their masses through the inclusion of properties such as spontaneous and explicit (for pseudoscalar mesons) symmetry breaking \cite{fortyfour}. 
In the CQMF model, the effective Lagrangian density accounting for various interaction terms  is given as \cite{fortyfour, fortyfive, fiftythree}
\begin{equation}
	{\cal L}_{{\rm effective}} \, = \, {\cal L}_{q0} \, + \, {\cal L}_{qm}
	\, +\, {\cal L}_{VV} \,+ \, {\cal L}_{\Sigma\Sigma} \,+ \, {\cal L}_{\chi SB}\, +
	\, {\cal L}_{cp}, \label{totallag}
\end{equation}
where $\, {\cal L}_{q0}=i\bar{q}\gamma^{\mu}\partial_{\mu}q \,$ represents the kinetic term for
quarks having zero mass, while $ {\cal L}_{qm}\,$ accounts for the interaction between quarks and mesons, which remains invariant when undergoing chiral $SU(3)$ transformations. $ {\cal L}_{qm}\,$ is expressed as \cite{fortyfour,fortysix,fiftythree,fiftyfour}
\begin{equation}\label{eq:second}
	\begin{aligned}
		\mathcal{L}_{q m} & =g_{\mathrm{s}}\left(\bar{\Psi}_R {M_n}^{\dagger} \Psi_L+\bar{\Psi}_L {M_n} \Psi_R\right)  -g_v\left(\bar{\Psi}_L \gamma^\mu L_\mu \Psi_L+\bar{\Psi}_R \gamma^\mu r_\mu \Psi_R\right) \\
		& =\frac{g_{\mathrm{s}}}{\sqrt{2}} \bar{\Psi}\left(\mathrm{i} \gamma^5 \sum_{b=0}^8 p_b \lambda_b+\sum_{b=0}^8 s_b \lambda_b\right) \Psi  -\frac{g_v}{2 \sqrt{2}} \bar{\Psi}\left(\gamma^\mu \sum_{b=0}^8 v_\mu^b \lambda_b-\gamma^\mu \gamma^5 \sum_{b=0}^8 a_\mu^b \lambda_b\right) \Psi.
	\end{aligned}
\end{equation}
In \cref{eq:second}, $g_{\mathrm{s}}$ and $g_v$ represent scalar and vector coupling constants, respectively and $\Psi=$ 
$ \begin{pmatrix}
	u \\ 
	d  \\
	s
\end{pmatrix}$. The relationships between these coupling constants are given as \cite{fortysix} 
\begin{align}
	\frac{g_s}{\sqrt{2}}
	= &g_{\delta}^u = -g_{\delta}^d = g_\sigma^u = g_\sigma^d  =
	\frac{1}{\sqrt{2}}g_\zeta^s, \label{relation1}
	~~~~~g_{\delta}^s = g_\sigma^s = g_\zeta^u = g_\zeta^d = 0 \, ,\\
\frac{g_v}{2\sqrt{2}}
	= 	& g_{\rho}^u = -g_{\rho}^d = g_\omega^u = g_\omega^d=\frac{g_{\phi}^s}{\sqrt{2}},
	~~~~~~~g_\omega^s = g_{\rho}^s  = g_{\phi}^u = g_{\phi}^d = 0. \label{relation2}
\end{align}
It is important to note that spin-0 scalar, $\Sigma$ and pseudo-scalar meson nonets  $\Pi$, are expressed as \cite{fortyfour,fortysix}
\begin{align}
	M_n\left({M_n}^{\dagger}\right)=\Sigma \pm i \Pi=\frac{1}{\sqrt{2}} \sum_{b=0}^8\left(s^b \pm i \pi^b\right) \lambda^b,
\end{align}
where $s^b$ and $\pi^b$ represent scalar and pseudo-scalar meson nonents, respectively, while $\lambda^b$ represents Gell-Mann matrices with $\lambda^0=\sqrt{\frac{2}{3}} I$. The spin-1 mesons can also be defined by the relation
\begin{align}
	L_\mu\left(r_\mu\right)=\frac{1}{2}\left(V_\mu \pm A_\mu\right)=\frac{1}{2 \sqrt{2}} \sum_{b=0}^8\left(v_\mu^b \pm a_\mu^b\right) \lambda^b .
\end{align}

In the above expression, $v_\mu^b$ and $a_\mu^b$ are  vector and pseudovector meson nonents, respectively. The expressions representing scalar and vector meson nonets are
\begin{align}
	\Sigma=\frac{1}{\sqrt{2}} \sum_{b=0}^8 s_b \lambda^b=\left(\begin{array}{lcc}
		\frac{1}{\sqrt{2}}\left(\delta^0+\sigma\right) & \delta^{+} & \kappa^{*+} \\
		\delta^{-} & \frac{1}{\sqrt{2}}\left(\sigma-\delta^0\right) & \kappa^{* 0} \\
		\kappa^{*-} & \bar{\kappa}^{* 0} & \zeta
	\end{array}\right),
\end{align}
and
\begin{align}
	V_\mu=\frac{1}{\sqrt{2}} \sum_{b=0}^8 v_\mu^b \lambda^b=\left(\begin{array}{lcc}
		\frac{1}{\sqrt{2}}\left(\omega_\mu+\rho_\mu^0\right) & \rho_\mu^{+} & K_\mu^{*+} \\
		\rho_\mu^{-} & \frac{1}{\sqrt{2}}\left(\omega_\mu-\rho_\mu^0\right) & K_\mu^{* 0} \\
		K_\mu^{*-} & \bar{K}_\mu^{* 0} & \phi_\mu
	\end{array}\right),
\end{align}
respectively. 

The chiral invariant term ${\cal L}_{VV}$ in \cref{totallag}, incorporates the mass term for the vector mesons as well as higher order self interactions and is written as \cite{fortyfour} 
\begin{equation}
	\begin{aligned}
		{\cal L}_{VV}=\frac{1}{2} m_V^2 \frac{\chi^2}{\chi_0^2} \operatorname{Tr} V_\mu V^\mu+2 g_4 \operatorname{Tr}\left(V_\mu V^\mu\right)^2.
		\label{lvec}
	\end{aligned}
\end{equation}
 The expression for ${\cal L}_{VV}$ under mean field approximation comes out to be \cite{fortyfour,fiftythree,fiftyfour}
\begin{align}
	{\cal L}_{VV}=\frac{1}{2} \, \frac{\chi^2}{\chi_0^2} \left(
	m_\omega^2\omega^2+m_\rho^2\rho^2+m_\phi^2\phi^2\right)+g_4\left(\omega^4+6\omega^2\rho^2+\rho^4+2\phi^4\right).
	\label{vector}
\end{align}

For the scalar mesons, their self interaction terms are written as (fourth term in \cref{totallag}) 
\begin{align}
	{\cal L}_{\Sigma\Sigma} =& -\frac{1}{2} \, k_0\chi^2
	\left(\delta^2+\zeta^2+\sigma^2\right)+k_1 \left(\delta^2+\zeta^2+\sigma^2\right)^2
	+k_2\left(3\sigma^2\delta^2+\frac{\delta^4}{2} +\frac{\sigma^4}{2}+\zeta^4\right)\nonumber \\ 
	&+k_3\chi\left(\sigma^2-\delta^2\right)\zeta 
	-k_4\chi^4-\frac14\chi^4 {\rm ln}\frac{\chi^4}{\chi_0^4} +
	\frac{\xi}
	3\chi^4 {\rm ln}\left(\left(\frac{\left(\sigma^2-\delta^2\right)\zeta}{\sigma_0^2\zeta_0}\right)\left(\frac{\chi^3}{\chi_0^3}\right)\right). \label{scalar0}
\end{align}

The parameters $k_{i=0,1,2,3,4}$ appearing in the above Lagrangian density are calculated using masses of $\pi$ and $K$ as well as 
the average of the $\eta$ and $\eta^{'}$ meson masses. The expectation values of scalar meson fields $\sigma$ and $\zeta$, in free space i.e. $\sigma_0$ and $\zeta_0$, can be expressed with pion and kaon leptonic decay constants $f_\pi$ and $f_K$, with the expressions
\begin{align}
	\sigma_0= -f_{\pi} ~~{\rm and}~~~~  \zeta_0= \frac{1}{\sqrt{2}}\left( f_{\pi}-2f_{K}\right),
\end{align} 
where $f_{\pi}=92.8$ MeV and $f_{K}=115$ MeV. 
In absence of nuclear matter, the dilaton field $\chi_0=254.6$ MeV and the coupling constant $g_4=37.4$, have been obtained from a reasonable fit of effective nucleon mass \cite{fortyfour,fiftythree,fiftyfour}. The concluding three terms of 
\cref{scalar0} incorporates the trace anomaly property of QCD, which leads to the trace of energy momentum tensor being proportional to the dilaton field $\chi$ raised to the power of 4 \cite{fortynine}.
The explicit chiral symmetry breaking term
 ${\cal L}_{\chi SB}$ in \cref{totallag} is expressed through the relation
\begin{equation}\label{L_SB}
	{\cal L}_{\chi SB}=\frac{\chi^2}{\chi_0^2}\left[m_\pi^2f_\pi\sigma +
	\left(
	\sqrt{2} \, m_K^2f_K-\frac{m_\pi^2}{\sqrt{2}} f_\pi\right)\zeta\right].
\end{equation}
The above expression accounts for 
the non-zero masses of pseudoscalar mesons and agrees well with the partial conserved axial-vector current relations of $\pi$ and $K$ mesons \cite{fortyfour,fortysix,fortyseven}. Unlike the models which uses the concept of spherical-bag boundary, like MIT bag Model, the CQMF model treats the constituent quarks to be moving inside the baryons forming a bounded system, by virtue of a confining potential $\chi_c$. The use of an average confining potential with quarks moving around the baryon core, has been fruitful in improving the static baryonic properties like magnetic moment of proton \cite{Barik1985}. The last term in the Lagrangian density, in \cref{totallag}, incorporates this as
\begin{align}
	{\cal L}_{cp} = -  \bar \psi \chi_c \psi,
\end{align}
and the confining potential $\chi_c$, is expressed as   
\begin{align}
	\chi_{c}(r)=\frac14 k_{c} \, r^2(1+\gamma^0) \,.   \label{potential}
\end{align}
In \cref{potential} above the coupling constant $k_c$ is considered to be $98 \, \text{MeV}. \text{fm}^{-2}$.

In light of  mean field approximation, we are to investigate the asymmetric strange matter properties at finite temperature and density, we are using \cite{fortyfour}. In the presence of meson mean field, we express the Dirac equation for the quark field $\Psi_{qi}$ as 
\begin{equation}
	\left[-i\vec{\alpha}\cdot\vec{\nabla}+\beta m_q^*+\chi_c(r)\right]
	\Psi_{qi}=e_q^*\Psi_{qi}. \label{Dirac}
\end{equation}
Here the quark $q$ is in a baryon of type $i$
and $\vec{\alpha}$\,, $\beta$\, are the Dirac matrices.
By definition, the in-medium constituent quark mass $m_{q}^*$ is
\begin{equation}
	m_q^*=-g_\sigma^q\sigma - g_\zeta^q\zeta - g_\delta^q I^{3q} \delta + m_{q0}. \label{qmass}
\end{equation}
In free space, the masses of constituent quarks $u$, $d$, $s$ 
are expressed in terms of $\sigma_0$ and $\zeta_0$, by the relation
\begin{align}
	\label{qvacmass}
	m_u=m_d=-\frac{g_s}{\sqrt{2}}\sigma_0=-g_{\sigma}^q \sigma_0.
\end{align}
The coupling constant $g_s$ and confining parameter $k_c$ are fitted to obtain binding energy of $-16.0$ MeV at nuclear saturation density, $\rho_0 = 0.16$ $\text{fm}^{-3}$ \cite{fortyfive}. This fit provides the value of effective mass of light quarks to be equal to $256$ MeV. 
Here $m_{q0}$ is zero for non-strange $u$ and $d$ quarks, while for $s$ quark $m_{q0}=\Delta m=77$ MeV.
In \cref{Dirac}, $e_q^*$ in the Dirac equation is the effective energy of  quarks   and  is expressed as  \cite{fortyfour}
\begin{equation}
	e_q^*= e_q - g_\omega^q\omega - g_\rho^q I^{3q}\rho -g_\phi^q\phi. \label{qenergy}
\end{equation}
The in-medium mass of a baryon is expressed as,
\begin{equation}
	M^{*}_i = \sqrt{E^{*2}_i - <p^{*2}_{icm}>}  ,
	\label{baryonmass}
\end{equation}
in terms of its in-medium energy $E^{*}_i$ and spurious center mass motion $p_{icm}$ \cite{Barik1985,Barik2013}.
  Spuirous center of mass momentum $p_{icm}$ corrections are added in expression for $M_{i}^{*}$ to further improve the overlooked
 independent quark motion inside the baryons.
 The explicit expression for $<p^{*2}_{icm}>$ being summation over average value of square of each constituent quark momentum ($<p^{*2}_{cm}>_q$) \cite{Barik1985}, i.e.,
\begin{equation}
	<p^{*2}_{icm}> = \sum_{q}<p^{*2}_{cm}>_q = \sum_{q} \frac{11 e_q^*+m_q^*}{6(3e_q^*+m_q^*)}(e_q^{*2}-m_q^{*2}).
\end{equation}
As can be seen from the above equation, spurious center of mass motion corrections are density and temperature dependent.
In Ref. \cite{fortyseven}, spurious center of mass motion corrections are considered
as independent of medium effects.
Further the in-medium energies of $i^{th}$ baryon can be related to the in-medium energies of quarks and number of \textit{q} type constituent quarks $n_{qi}$ in $i^{th}$ baryon by the expression
\begin{equation}
	E^{*}_i = \sum_qn_{qi}e_q^{*} + E_{ispin}.
	\label{eispin}
\end{equation}
Here, $E_{ispin}$ accounts for spin-spin interaction of constituent quarks and serves as a correction term in the effective energy of a baryon. The $E_{ispin}$ is computed so as to obtain the correct baryon masses at zero density.
Finally, the thermodynamic potential $\Omega$ describing the  isospin asymmetric strange hadronic matter at finite temperature and baryon density is written as \cite{fortyfour}
\begin{equation}
	\Omega=-\frac{k_B T}{(2 \pi)^3} \sum_i \gamma_i \int_0^{\infty} \left\{\ln \left(e^{-\left[E_i^*(k)-\nu_i^*\right] / k_B T}+1\right)+\ln \left(e^{-\left[E_i^*(k)+\nu_i^*\right] / k_B T}+1\right)\right\}d^3 k-\mathcal{L^\prime}-\mathcal{V}_{\text {vac }},
	\label{eq:thermo}
\end{equation}
where the mesonic Lagrangian is $\mathcal{L^\prime} = \mathcal{L}_{VV} + \mathcal{L}_{\Sigma\Sigma} + \mathcal{L}_{\chi SB}$ and we have $E^*_{i}(k)=\sqrt{M_i^{* 2}+k^2}$. In the above equation, the degeneracy factor is $\gamma_i = 2$ for baryons and the summation is over hyperons and nucleons. In \cref{eq:thermo}, the effective chemical potential
$\nu_i^*$ is expressed in terms of the usual baryon chemical potential $\mu_i$ as \cite{fortyfour,fiftythree,fiftyfour,sixtytwo}
\begin{equation}
	\nu_i^*=\mu_i-g_\phi^i \phi-g_\omega^i \omega-g_\rho^i I^{3 i} \rho .
	\label{eq:chem_potential}
\end{equation}
From \cref{eq:thermo} the equations of motion of scalar and vector meson fields are calculated by minimizing the effective thermodynamic potential as
\begin{equation}
	\frac{\partial \Omega}{\partial \sigma}=\frac{\partial \Omega}{\partial \zeta}=\frac{\partial \Omega}{\partial \chi}=\frac{\partial \Omega}{\partial \delta}=\frac{\partial \Omega}{\partial \omega}=\frac{\partial \Omega}{\partial \rho}=\frac{\partial \Omega}{\partial \phi}=0 .
	\label{eq:eom}
\end{equation}
\Cref{eq:eom} is solved for finite values of temperature and densities of the medium. 
Note that the confining potential $\chi_c$ contribute to the effective energy $e_q^{*}$ and hence, masses $m_q^{*}$ of quarks through the solution of Dirac equation. The effective quark energies are used in calculations of baryon energy which further enter into the coupled system of equations for 
scalar and vector fields through the definition of scalar and vector density of baryons \cite{fortyfour}.
In order to account for isospin asymmetry in the hadronic medium a parameter $i_a =-\frac{\Sigma_i I_{3 i} \rho_i}{\rho_B}$ is introduced, where $I_{3 i}$ represents the third component of isopin and $\rho_B$ is baryonic density of the medium \cite{fiftyfour}.
To take into account hyperons in the medium,  we define
strangeness fraction $f_s=\frac{\Sigma_i\left|s_i\right| \rho_i}{\rho_B}$,  here $|s_i|$ denotes the number of $s$ quarks in $i^{th}$ baryon.
Note that in the present manuscript,  the difference in the $\frac{1}{2}^{-}$ baryon resonances 
  and positive parity octet baryons comes through the fitting of $E_{ispin}$ value to their respective vacuum masses.
 The calculations of effective masses of $\frac{1}{2}^{-}$ baryon resonances  proceed in two steps: 
 first we solve the system of equations corresponding to the scalar ($\sigma, \zeta$ and $\delta$) and vector fields ($\omega, \rho$ and $\phi$) in the strange hadronic medium (considering only positive parity octet baryons and utilizing Eqs. (\ref{baryonmass}) and (\ref{eispin}) for in-medium effective masses and energies) at finite density and temperature (see Eq. (\ref{eq:eom})).
This gives us the density and temperature dependent values of scalar and vector fields which are used 
 in the second step to obtain the effective masses of $\frac{1}{2}^{-}$ baryon resonances.
For this, first we use the vacuum  ($\rho_B = 0, T= 0, i_a =0, f_s =0$) values of fields calculated in first step, in Eqs. 
(\ref{baryonmass}) and (\ref{eispin} and fit $E_{ispin}$ to vacuum masses of $\frac{1}{2}^{-}$ baryon resonances.
Using these $E_{ispin}$ values 
and density and temperature dependent values of scalar and vector fields calculated in the first step (strange matter), the in-medium masses of  $\frac{1}{2}^{-}$ baryon resonances are calculated using Eqs. (\ref{baryonmass}) in the strange hadronic medium.

We have \cref{baryonmass} which addresses our aim to compute the effective baryon masses, by taking into account the effective constituent quark masses, baryon density and the spurious center of mass motion. In the upcoming section we shift the trajectory of our discussion towards the calculation of effective baryonic magnetic moments, by incorporating the effective in-medium masses of constituent quarks and the baryons under study.

\subsection{Magnetic Moment of ${\frac{1}{2}}^-$ baryon resonances}
\label{sec:magnetic}

The formalism of chiral constituent quark model ($\chi \mathrm{CQM}$) \cite{thirtyone}, has been implemented to calculate the magnetic moment of the baryons through the following fluctuation process \cite{eight,fiftyfive}
\begin{equation}
	q^{\uparrow \downarrow} \rightarrow \mathrm{NGB}+q^{\prime \downarrow \uparrow} \rightarrow {\left(q \overline{q}^{\prime} \right)+q^{\prime \downarrow \uparrow}},
	\label{eq: fluctuation}
\end{equation}
where the emitted Nambu-Goldstone boson (NGB) furthermore splits into a $q \overline{q}^{\prime}$ pair, and $q \overline{q}^{\prime}+q^{\prime}$ constitute the sea quarks \cite{nineteen,twenty,twentyone,twentytwo,twentythree}. In $\chi \mathrm{CQM}$ model the massless confined quarks obtain its mass by virtue of spontaneous chiral symmetry breaking \cite{twelve,fiftysix}. The coupling between the quark and NGBs is described by the effective interaction Lagrangian
\begin{equation}
	\mathcal{L}=g_8 \overline{\mathbf{q}}\left(\Phi^{\prime}\right) \mathbf{q}=g_8 \overline{\mathbf{q}}\left(\zeta^{\prime} \frac{\eta^{\prime}}{\sqrt{3}} I+\Phi\right) \mathbf{q},
	\label{eq: lagrangian_GB}
\end{equation}
where $\zeta^{\prime}=g_1 / g_8,$ where $g_1$ and $g_8$ denote the coupling constants for the singlet and octet NGBs respectively and $I$ is an order 3 identity matrix.
This quark-NGB coupling represented in this Lagrangian is quite feeble and hence, the fluctuation in \cref{eq: fluctuation} is practically taken to be a minor perturbation. The spin-flip process is an outcome of the emission of NGB.  

The NGB matrix is defined by
\begin{equation}
	\begin{aligned}
		& \Phi^{\prime}=\left(\begin{array}{ccc}
			\frac{\pi^0}{\sqrt{2}}+\beta \frac{\eta}{\sqrt{6}}+\zeta^{\prime} \frac{\eta^{\prime}}{\sqrt{3}} & \pi^{+} & \alpha K^{+} \\
			\pi^{-} & -\frac{\pi^0}{\sqrt{2}}+\beta \frac{\eta}{\sqrt{6}}+\zeta^{\prime} \frac{\eta^{\prime}}{\sqrt{3}} & \alpha K^0 \\
			\alpha K^{-} & \alpha \bar{K}^0 & -\beta \frac{2 \eta}{\sqrt{6}}+\zeta^{\prime} \frac{\eta^{\prime}}{\sqrt{3}}
		\end{array}\right) \\
		& \text { and } \\
		& q=\left(\begin{array}{l}
			u \\
			d \\
			s
		\end{array}\right) . \\
		&
	\end{aligned}
\end{equation}
The $SU(3)$ chiral symmetry breaking is introduced in $\chi \mathrm{CQM}$ by treating the masses of NGBs as non-degenerate
and considering that $m_s>m_{u, d}$ \cite{fiftyseven}. 
The parameter for the transition probability of chiral fluctuations: $u(d) \rightarrow$ $d(u)+\pi^{+(-)}$ is denoted by $a(=\left|g_{8}\right|^2)$, $u(d) \rightarrow s+K^{-(0)}$ by $a \alpha^2$, $u(d, s) \rightarrow$ $u(d, s)+\eta$ by $a \beta^2$ and $ u(d, s) \rightarrow u(d, s)+\eta^{\prime}$ is represented by $a \zeta^{\prime 2}$ \cite{nineteen,twenty,twentyone,twentytwo,twentythree}.
By thoroughly analyzing the spin and flavor distribution functions of proton, the best fitted set of parameters crucial to our calculations are \cite{nineteen,twentythree,twentyfive,twentyeight,twentynine,thirty, thirtyone} : $a =0.12, \alpha = \beta =0.45, \zeta^\prime = -0.15$.

The effective baryonic magnetic moment $\mu^*_{B}$ is a summation of the contributions due to spin polarization $\mu^{*S}_{B}$ and the angular momentum polarization $\mu^{*L}_{B}$, expressed as \cite{fiftythree,fiftyfour}
\begin{align}
	\mu^*_{B}= \mu^{*S}_{B} + \mu^{*L}_{B}.
	\label{magtotal}
\end{align}

Following the Cheng Li mechanism \cite{ten}, the calculation of total baryonic magnetic moment involves the valence and sea quark contributions and the orbital angular momentum of sea quarks. The spin part in \cref{magtotal}, comprises of these individual contributions and is given by
\begin{align}
	\mu^{*S}_B = \mu^{*S}_{val} + \mu^{*S}_{sea} + \mu^{*S}_{orbit}.
\end{align}

The $\mu^{*S}_{val}$, of a baryon, can be derived by calculating the magnetic moments for the physical eigenstates of the orbital angular momentum $L = 1$ ($P-$wave) negative-parity resonances \cite{thirtyone}.
For each baryon, the contribution of sea quark spin polarization in spin part of the total magnetic moment is obtained by substituting for each valence quark
\begin{equation}
	q^{\uparrow(\downarrow)} \rightarrow-P_{[q, G B] q^{\uparrow(\downarrow)}}+\left|\psi\left(q^{\uparrow(\downarrow)}\right)\right|^2.
	\label{sea_spin}
\end{equation}
In the above expression, the emission probability of NGBs from a quark $q^{\uparrow(\downarrow)}$ is $P_{[q, NGB]}$ which can be fetched from ref. \cite{fiftyeight},
and $\left|\psi\left(q^{\uparrow(\downarrow)}\right)\right|^2$ represents the probability of transforming a $q^{\uparrow(\downarrow)}$ quark 
\cite{nineteen,twenty,twentyone,twentytwo,twentythree,fiftythree,fiftyfour}.
The orbital angular momentum contribution due to sea quarks, is given by \cite{ten,fourteen,twentyeight,twentynine,thirty,thirtyone}
\begin{equation}
	\mu^*\left(q^{\uparrow} \rightarrow q^{\prime \downarrow}\right)=\frac{e_q}{2 m^*_q}\left\langle l_q\right\rangle+\frac{e_q-e_{q\prime}}{2 M_{\mathrm{NGB}}}\left\langle l_{\mathrm{NGB}}\right\rangle,
	\label{eq:orb_angm}
\end{equation}
where $\left<l_q\right>=\frac{M_{\mathrm{NGB}}}{M_{\mathrm{NGB}}+m^*_q}$ and $\left<l_{\mathrm{NGB}}\right>=\frac{m^*_q}{M_{\mathrm{NGB}}+m^*_q}$. The quantities $\left(l_q,m^*_q\right)$ and $\left(l_{\mathrm{NGB}}, M_{\mathrm{NGB}}\right)$ stands for the mass and orbital angular moment of constituent quarks and NGBs, respectively. The probability of each process to occur, allows us to compute the magnetic moment due to all possible transitions for a given valence quark. 

The orbital moments of the constituent quarks can also be written using the $\chi \mathrm{CQM}$ parameters $(a, \alpha, \beta, \zeta^{\prime})$, NGB masses $\left(M_{\pi}, M_{\eta\prime}, M_K, M_{\eta}\right)$ and effective quark masses $\left(m^*_u, m^*_d, m^*_s\right)$, respectively as \cite{thirtyone,fiftythree,fiftyfour}:
  
\begin{align}
	{\left[\mu^*\left(u_{\uparrow} \rightarrow\right)\right]=} & a\left[\frac{3}{2}\frac{m_u^{*2}}{M_\pi\left(m_u^{*}+M_\pi\right)}-\frac{\alpha^2}{2}\frac{\left(M_K^2-3 m_u^{*2}\right)}{M_K\left(M_K+m_u^{*}\right)}\right. +\frac{\beta^2}{6}\frac{M_\eta}{6\left(M_\eta+m_u^{*}\right)} \notag \\& 
	\left.+\frac{\zeta^{\prime2}}{3}\frac{M_{\eta^{\prime}}}{\left(m_u^{*}+M_{\eta^{\prime}}\right)}\right] \mu^*_u, \notag \\
	%
	{\left[\mu^*\left(d_{\uparrow} \rightarrow\right)\right]=} & -2a \left[\frac{3}{4}\frac{\left(M_\pi^2-2 m_d^{*2}\right)}{M_\pi\left(m^*_d+M_\pi\right)}-\frac{\alpha^2}{2}\frac{M_K}{\left(m_d^*+M_K\right)} -\frac{\beta^2}{12}\frac{M_\eta}{\left(m_d^*+M_\eta\right)} \right. \label{orbmom_tag} \\& 
	\left.-\frac{\zeta^{\prime2}}{6}\frac{M_{\eta^{\prime}}}{\left(m_d^*+M_{\eta^{\prime}}\right)}\right] \mu^*_d, \notag \\ 
	%
	{\left[\mu^*\left(s_{\uparrow} \rightarrow\right)\right]=} & -2a \left[\frac{\alpha^2}{2}\frac{\left(M_K^2-3 m_s^{*2}\right)}{M_K\left(m_s^*+M_K\right)}  -\frac{\beta^2}{3}\frac{M_\eta}{\left(m_s^*+M_\eta\right)}  -\frac{\zeta^{\prime2}}{6}\frac{M_{\eta^{\prime}}}{\left(m_s^*+M_{\eta^{\prime}}\right)}  \right] \mu_s^*. \notag 
\end{align}
 
In the present study, we have included the effect of relativistic quark confinement and have calculated the mass adjusted in-medium magnetic moments of its constituent quarks by the expression \cite{fiftyfive}
\begin{equation}
	\mu_u^*=2\left(1-\frac{\Delta M_B}{M_B^*}\right), \quad \mu_d^*=-\frac{1}{2} \mu_u^*, \quad  \mu_s^*=-\frac{m_u^*}{m_s^*}\left(1-\frac{\Delta M_B}{M_B^*}\right).
	\label{quark_mm}
\end{equation}
In the above relations, $\Delta M_B = M_{B,vac} - M_B^*$. Here, $M_{B,vac}$ and $M_B^*$ denotes the vacuum and in-medium mass of a given baryon, respectively. 
As discussed in the introduction of present manuscript, in Ref. \cite{thirtyone} the impact of relativistic quark confinement was not considered and the simplified expression $\mu_q=\frac{e_q}{2 m_q}, q=u,d,s$ was used to calculate the magnetic moment of constituent quarks in the free space.
To account for the contributions of orbital angular momentum polarizations ($\mu^{*L}_B$) towards the total baryonic magnetic moment, we have calculated the individual contributions from the valence quarks and quark sea as
\begin{equation}
	\mu_B^{*L}=\mu_{\mathrm{val}}^{*L}+\mu_{\mathrm{sea}}^{*L}.
\end{equation}
In this equation, $\mu_{\mathrm{val}}^{*L}$ and $\mu_{\mathrm{sea}}^{*L}$ represent the effective contributions of the valence and sea quarks towards the baryonic magnetic moment component, $\mu^{*L}_B$, due to the orbital angular momentum polarization. The details of the valence quark calculations has been discussed in ref. \cite{thirtyone}. In the current section we will also describe the expressions used in the calculation of sea quark spin polarization in a baryon, which are obtained by substituting the third component of the orbital angular momentum of sea quarks in place of each valence quark as,
\begin{equation}
	q^{( \pm 1)} \rightarrow-T_{[q, NGB]} q^{( \pm 1)}+\left|\psi\left(q^{( \pm 1)}\right)\right|^2.\hspace{5.3cm}
	\label{gdd}
\end{equation}
In \cref{gdd}, $T_{[q, NGB]}$ refers to the emission probability of a Nambu-Goldstone bosons (NGB) from a quark $q^{( \pm 1)}$  and $\left|\psi\left(q^{( \pm 1)}\right)\right|^2$ represents the transformation probability of a $q^{( \pm 1)}$ quark into two other quarks of $SU(3)$ model, as given in Ref. \cite{thirtyone}.

Using the probability of transition of a constituent quark ($q^{ \pm 1}$) into all other quarks ($u, d$ and $s$) accompanied by NGBs \cite{fiftyeight}, we can easily obtain the $T_{[q, NGB]}$ for each constituent quark. The contributed values from the valence and sea quark polarizations of the orbital angular momentum part of the magnetic moment of the baryons, can thus be calculated easily. 
Working on the framework as described in this section, the magnetic moments of the $\frac{1}{2}^-$ octet baryon resonances $p^*, n^*, \Sigma^{*+}, \Sigma^{*-}, \Sigma^{*0}, \Xi^{*-}$ and $\Xi^{*0}$ as well as the low-lying $\frac{1}{2}^-$ $N^*$ resonances $S_{11}^{+}(1650)$, and $S_{11}^0(1650)$ have been calculated within the $\chi \mathrm{CQM}$ framework.

\section{Results \& Discussion} \label{sec:results}
We now discuss numerical results on the in-medium masses (\cref{sec_quark_masses}) and magnetic moments (\cref{sec_deculet_moments}) of spin $\frac{1}{2}^{-}$ octet baryon resonances. Various parameters used in the current manuscript are listed in Table \ref{tab:array-table}. The effective baryon masses have been listed in Tables \ref{tab:masses_T0} and \ref{tab:masses_T100}, so as to facilitate a discussion on the mass modifications. As an outcome of the calculations discussed in the previous sections, the vacuum and in-medium magnetic moments of $\frac{1}{2}^-$ baryon resonances have been illustrated in Tables \ref{table:mub_eta0fs0} to 
\ref{table:mub_eta5fs3}. For the sake of comparison with the baryonic magnetic moments obtained from earlier studies, we have tabulated the values calculated through different models in Table \ref{table:table5} for $\rho_B = 0$.
As discussed in previous section, in our present work quark masses and energies are defined in terms of scalar and vector fields. The  values of the confining parameter $k_c$ and couplings are fitted to binding energy value of $-16$ MeV at the nuclear saturation density. With this approach, at $\rho_B = 0$, the constituent quark masses of $u$ and $d$ quarks are obtained to be 256 MeV while  for s-quark it is 477 MeV. The difference in the constituent quark masses in present manuscript compared to Ref. \cite{thirtyone} (for $u/d$ quarks 330 MeV and for $s$ quark 490 MeV), leads to different values for the magnetic moments of baryons in the free space as can be seen in Table \ref{table:table5}.

\subsection{In-medium masses of $\frac{1}{2}^-$ baryon resonances}
\label{sec_quark_masses}
\begin{figure}[!ht] 
	\hspace*{-1cm}
	\includegraphics[width=\textwidth, height=16 cm]
	{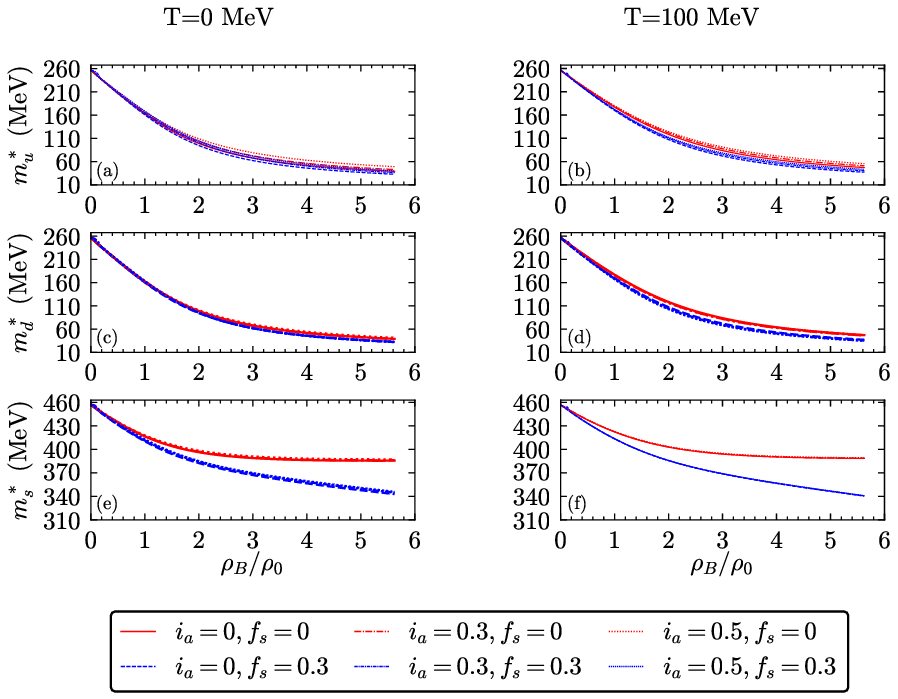}\hfill
    \caption{In-medium masses of constituent quarks $u, d$ and $s$, at varying medium density $\rho_B$ (in $\rho_0$ units).}
	\label{fig_massquarks}
\end{figure}
As the in-medium baryon masses is determined using the medium modified masses of constituent quarks, we first discuss the impact of density and temperature on the effective quark masses. 
In \cref{fig_massquarks}, the constituent quark masses, $m^*_u, m^*_d$ and $m^*_s$, are shown to vary with the relative hadronic matter density $\rho_B/\rho_0$, at $T = 0$ MeV (on the left) and $T = 100$ MeV (on the right)(where nuclear saturation density $\rho_0 = 0.16$ $\text{fm}^{-3}$). It also illustrates the in-medium mass modifications of quarks, subject to a medium with finite isospin asymmetry ($i_a$) and strangeness fraction ($f_s$). 
From \cref{fig_massquarks} it is observed that as we move from vacuum ($\rho_B = 0$) to denser hadronic medium, the quark masses decrease monotonously. For symmetric nuclear matter at temperature $T = 0$ MeV, the effective masses of light quarks decreases by 71.14\% 
and that of $s$ quark decreases by 14.65\%, 
at density value of $\rho_B =3\rho_0$. As observed, the effects of density on the $s$ quark is significantly lower than that on the light quarks, which is an implication to the negligible coupling of $s$ quark with $\sigma$ field \cite{fiftyfour,sixtytwo}. By introducing finite strangeness in the symmetric nuclear medium, at $f_s = 0.3$ and $T = 0$ MeV, with an increase in baryon density there is an appreciable drop in $m^*_s$, relative to that in symmetric nuclear matter, while in the light quarks only slight variations in the $m^*_u$ and $m^*_d$ values are observed.
It is also observed that with the rise in temperature at $T = 100$ MeV the effective masses of $u$ and $d$ quarks rises by 8.91 \% at $\rho_B =\rho_0$ and the $s$ quark mass elevates by just 1.38 \%. 
Thus an increase in the medium temperature causes a rise in the masses of constituent quarks, at the same density. 
Keeping the medium density affixed, a rise in isospin asymmetry parameter leads to a surge in-medium masses of constituent quarks. 
A study on the basis of an extended NJL model has previously demonstrated that while the in-medium masses of light quarks approach their current quark masses, the $s$ quark condensate remains large even at densities close to $\rho_B \approx 1$ $\text{fm}^{-3}$ \cite{xia}.

\begin{figure}[!ht] 
	\hspace*{-1cm}
	\includegraphics
	[width=\textwidth, height=16 cm]{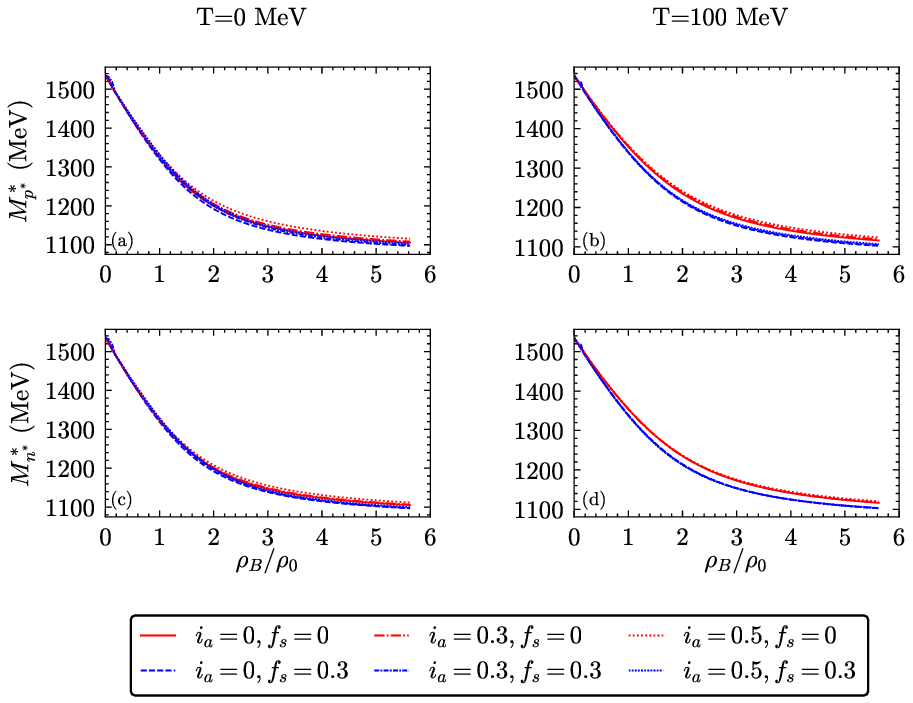}\hfill
	\caption{In-medium masses of $\frac{1}{2}^-$ octet baryon resonances $p^*$ and $n^*$, at varying medium density $\rho_B$ (in $\rho_0$ units).}
	\label{fig_mass_pn}
\end{figure}

\begin{figure}[!ht] 
	\hspace*{-0.9cm}
	\includegraphics[width=\textwidth, height=16 cm]
	{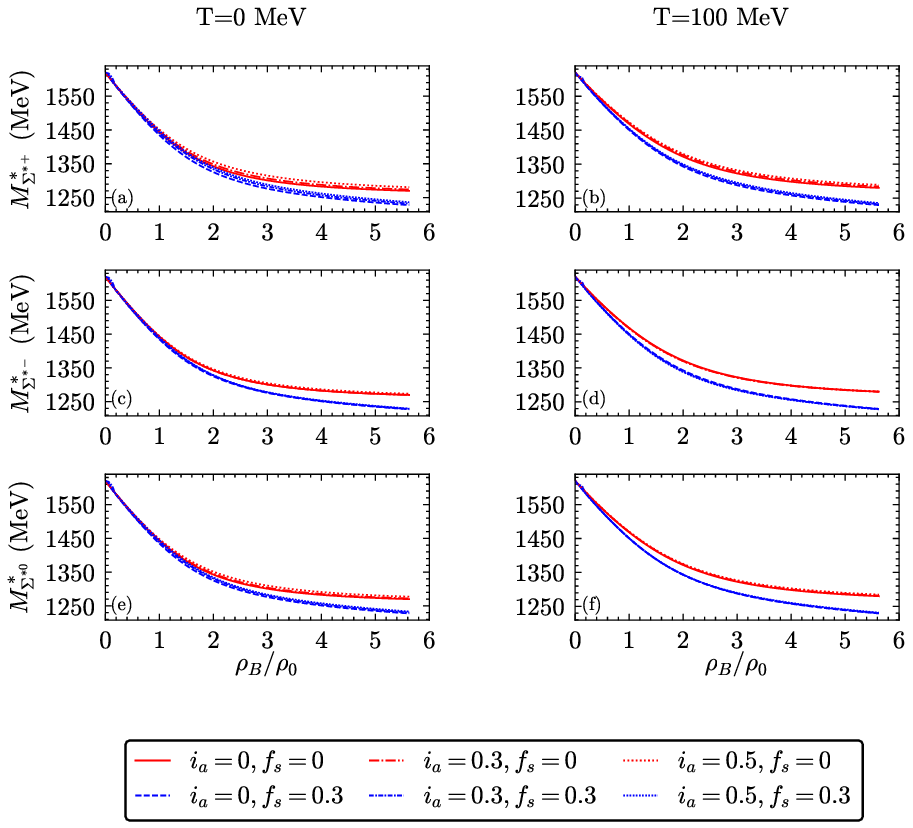}\hfill
	\caption{In-medium masses of $\frac{1}{2}^-$ octet baryon resonances $\Sigma^{*+}$, $\Sigma^{*-}$ and $\Sigma^{*0}$, at varying medium density $\rho_B$ (in $\rho_0$ units).}
	\label{fig_mass_sigma}
\end{figure}

\begin{figure}[!ht] 
	\hspace*{-1cm}
	\includegraphics[width=\textwidth, height=16 cm]
	{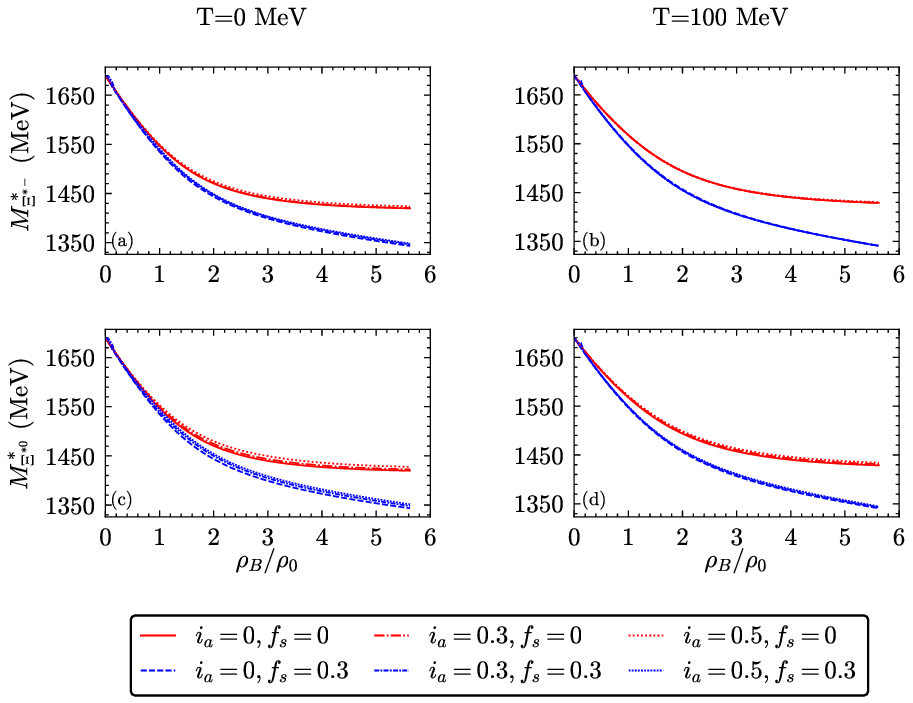}\hfill
		\caption{In-medium masses of $\frac{1}{2}^-$ octet baryon resonances $\Xi^{*-}$ and $\Xi^{*0}$, at varying medium density $\rho_B$ (in $\rho_0$ units).}
	\label{fig_mass_xi}
\end{figure}

\begin{figure}[!ht] 
	\hspace*{-1cm}
	\includegraphics
	[width=\textwidth, height=16 cm]
	{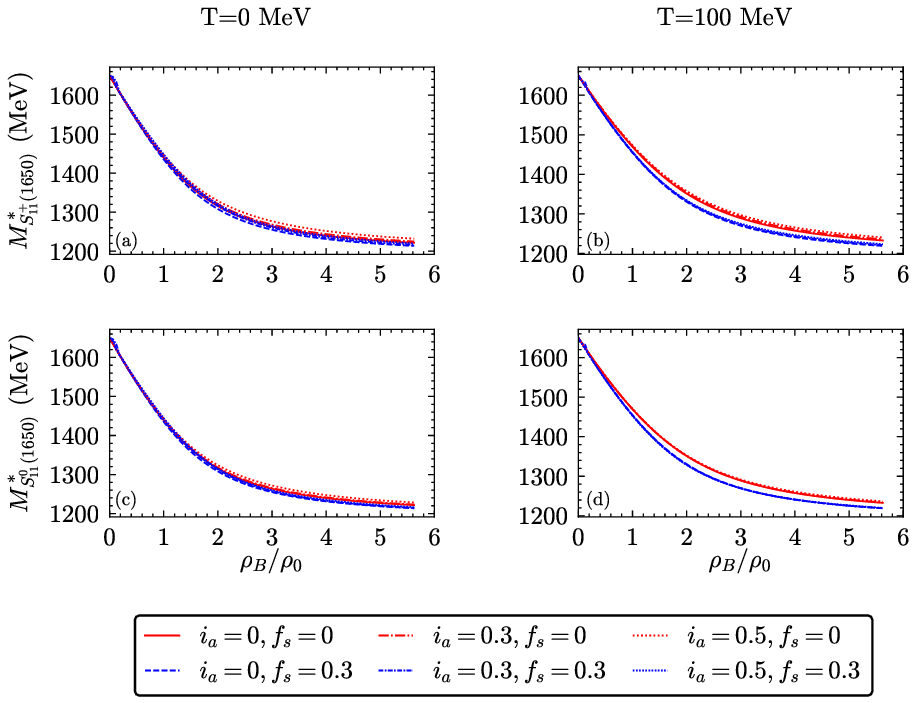}\hfill
	\caption{In-medium masses of $\frac{1}{2}^-$ $N^*$ resonances $S_{11}^{+}(1650)$ and $S_{11}^{0}(1650)$, at varying medium density $\rho_B$ (in $\rho_0$ units).}
	\label{fig_mass_s11_1650}
\end{figure}

In \cref{fig_mass_pn,fig_mass_sigma,fig_mass_xi} and \cref{fig_mass_s11_1650} we have shown the variations in effective baryonic masses, as a function of medium density, for $\frac{1}{2}^-$ octet baryon resonances and $N^*$ resonance $S^{+(0)}_{11} (1650)$, respectively. 
Using \cref{baryonmass} the in-medium masses of $\frac{1}{2}^-$ baryon resonances are computed in the presence of a hadronic matter, where the $E_{ispin}$ values are best fitted to procure the masses of $\frac{1}{2}^-$ baryon resonances in free space. A quick glance at the arguments and figures depicts, a monotonous decrease in the in-medium baryon masses as a function of density. The baryon masses in the medium fall rapidly in the lower density region up to $\rho_B = \rho_0$ and then saturate as the medium approaches large densities. As seen from \cref{fig_mass_pn}, when symmetric nuclear matter density rises upto $\rho_B =3 \rho_0$, the mass of $p^*$ resonance drops appreciably by a decrease of 25.27\%. At a density of $3 \rho_0$, with finite isospin asymmetry $i_a = 0.3$, the mass of $p^*$ is 1149.18 MeV which is a mild rise from 1147.02 MeV at $i_a = 0$. Taking into account a temperature rise from $T=0$ MeV to 100 MeV, we see distinctive baryonic mass curves for different strangeness fractions ($f_s = 0, 0.3$) of the medium. This exact same behaviour is seen to be replicated by the medium modified mass plots of $n^*$ as can be seen from \cref{fig_mass_pn}, which is quite trivial as well considering the light quark compositions of $p^*$ and $n^*$. 
In contrary to light quark content baryons, for symmetric nuclear matter the mass modifications incurred by $\Sigma^*$ resonances, as seen in \cref{fig_mass_sigma}, is 18.5\% and that of $\Xi^*$ resonances in \cref{fig_mass_xi} is 14.5\%. We can clearly state that, the density dependency on the baryons constituted by light quarks is much higher than that of hyperons, which is because $m^*_u$ and $m^*_d$ are more sensitive to density changes as compared to $m^*_s$. Additionally in hyperons, the in-medium effects are subject to the extent of $s$ quarks constituting a baryon. This is because as the number of constituent $s$ quarks increase, the in-medium effects on the constituent quark masses start having a dominating influence on the total mass of the baryon \cite{sixtytwo}.
In \cref{fig_mass_s11_1650}, the masses of $N^*$ resonances $S_{11}^+ (1650)$ and $S_{11}^0 (1650)$ varying as a function of medium density, have been shown. The extent of in-medium variations of the $N^*$ resonances is large compared to hyperons, like, in the case of $p^*$ and $n^*$ resonances. 
At a given value of $\rho_B$, with $i_a = 0, T = 0$ MeV, the influence of non-zero strangeness fraction, $f_s$, leads to a sizable decrease in the masses of baryons. Looking up the mass of $\Xi^{*-}$ from \cref{tab:masses_T0}, at $\rho_B =3 \rho_0, f_s = 0.3(0)$, we observe that the value decreases to 1399.14(1439.75) MeV. A cursory look at the plots of effective baryonic masses versus medium density suggests that, if the temperature of the medium increases to $T = 100$ MeV then the mass modification curves will have relatively less steeper slope than that at $T= 0$ MeV. For example, at $i_a= 0, f_s = 0(0.3)$ when the medium density is at $\rho_B =3\rho_0$, the effective mass of $\Sigma^{*+}$ being of 1300.88(1276.30) MeV, in \cref{tab:masses_T0}, at $T = 0$ MeV increases to 1322.58(1287.69) MeV, in \cref{tab:masses_T100}, at $T = 100$ MeV.

Owing to the strong coupling of $N^*$(1535) nucleon resonance with $\eta N$, there has been an earlier work that leverages $\eta$ mesic nuclei as a tool to map the reduction in-medium mass difference between nucleon $N$ and its chiral partner $N^*$, in chiral doublet model \cite{2002_mass}. In another study conducted using parity doublet model at finite baryonic densities of cold nuclear matter, the authors showcases the dependency of the vacuum mass of $N^*$ to determine the critical density and the order of the phase transition \cite{2006_mass}. For negative parity nucleons, the density effects on the in-medium baryonic mass turns out to be weaker than the impact on nucleon mass \cite{2016_mass}, which also gets reflected in \cref{fig_mass_pn}. In the context of hot hadronic matter, a study based on hadron resonance gas (HRG) sheds light on the temperature dependency of the effective masses of negative parity baryons \cite{2018_mass}. The observed temperature effects within $T = 100$ MeV, have been in good agreement with the findings in current work.
In the context of $\frac{1}{2}^+$ octet baryons, there has been studies on their in-medium mass modifications based on QMC \cite{hadron_mass}, modified QMC (MQMC) \cite{mqmc}, CQMF \cite{sixtytwo} and also for decuplet, low-lying strange, charm, and bottom baryons in QMC framework \cite{tsushima1,tsushima2}.


\subsection{In-medium magnetic moments of ${\frac{1}{2}}^-$ baryon resonances}
\label{sec_deculet_moments}

\begin{figure}[!ht] 
	\hspace*{-1cm}
	\includegraphics[width=\textwidth, height=16 cm]{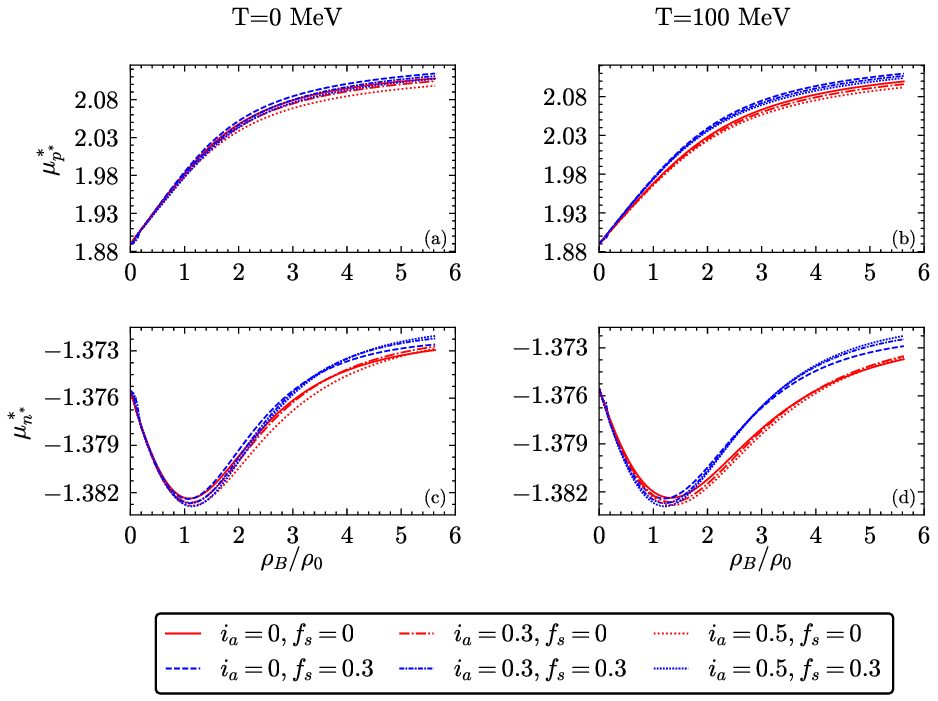}\hfill
	\caption{Variation of effective magnetic moments of $\frac{1}{2}^-$ octet baryon resonances $p^{*}$ and $n^{*}$ (in $\mu_N$ units) with baryon density $\rho_B$ (in units of $\rho_0$).}
	\label{fig_total_pn}
\end{figure} 
\begin{figure}[!ht]\hspace*{-1cm}
	\includegraphics[width=\textwidth, height=16 cm]{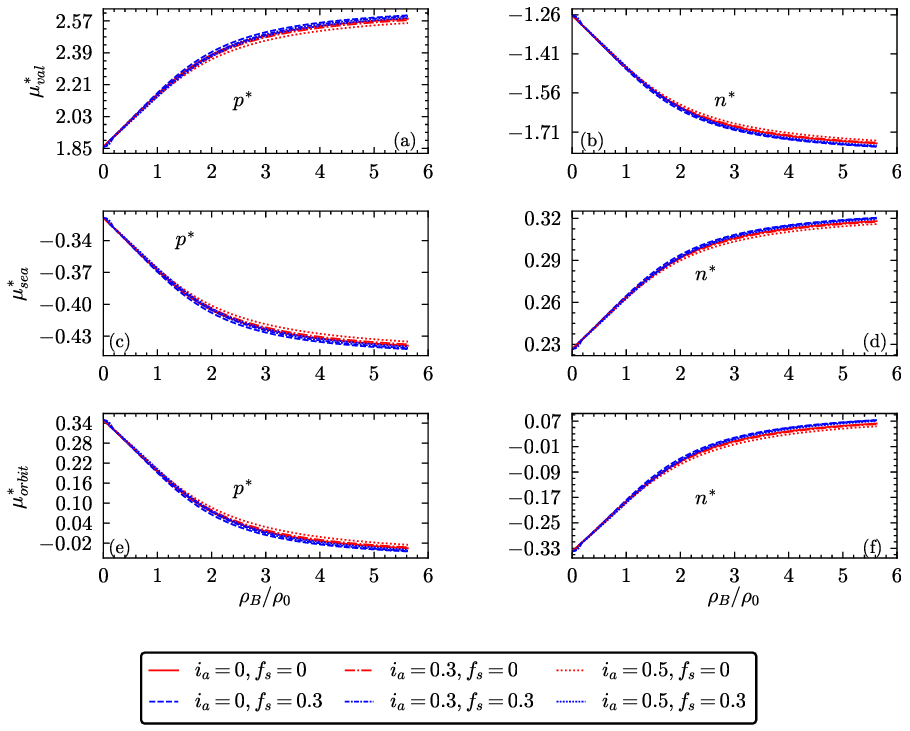}\hfill
	\caption{In-medium trends in explicit contributions to effective magnetic moments of $\frac{1}{2}^-$ octet baryon resonances $p^*$ and $n^*$ (in $\mu_N$ units), from valance quarks ($\mu_{val}^{*}$), sea quarks ($\mu_{sea}^{*}$) and orbital moment of sea quark ($\mu_{orbit}^{*}$) are presented.}
	\label{fig_contri_pn}
\end{figure} 
\begin{figure}[!ht] 
	\hspace*{-1.2cm}
	\includegraphics[width=\textwidth, height=16 cm]{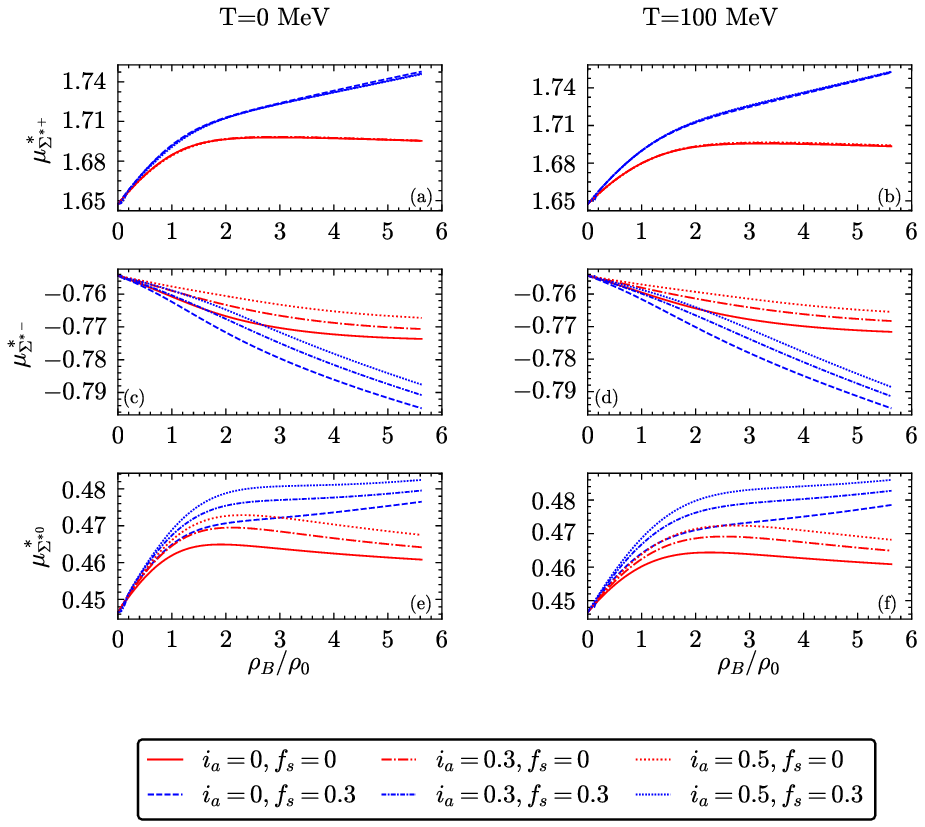}\hfill
		\caption{Variation of effective magnetic moments of $\frac{1}{2}^-$ octet baryon resonances $\Sigma^{*+}$, $\Sigma^{*-}$ and $\Sigma^{*0}$ (in $\mu_N$ units) with baryon density $\rho_B$ (in units of $\rho_0$).}
	\label{fig_total_sigma}
\end{figure}
\begin{figure}[!ht]
	\hspace*{-1.2cm}
	\includegraphics[width=\textwidth, height=16 cm]{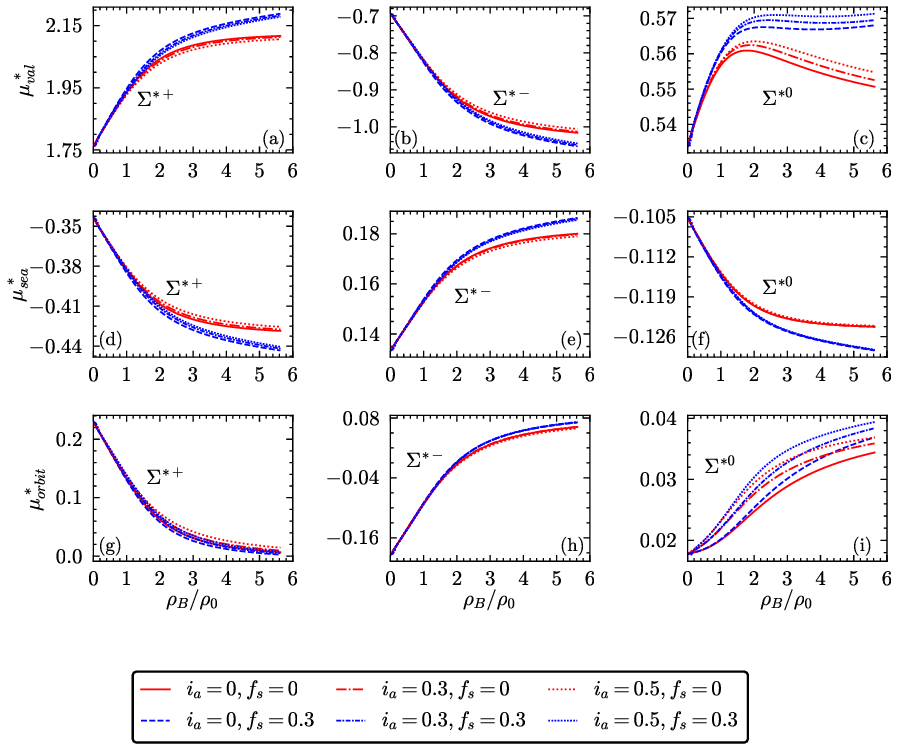}\hfill
	\caption{In-medium trends in explicit contributions to effective magnetic moments of $\frac{1}{2}^-$ octet baryon resonances $\Sigma^{*+}$, $\Sigma^{*-}$ and $\Sigma^{*0}$ (in $\mu_N$ units), from valance quarks ($\mu_{val}^{*}$), sea quarks ($\mu_{sea}^{*}$) and orbital moment of sea quark ($\mu_{orbit}^{*}$) are presented.}
	\label{fig_contri_sigma}
\end{figure}
\begin{figure}[!ht] 
	\hspace*{-1.2cm}
	\includegraphics[width=\textwidth, height=16 cm]{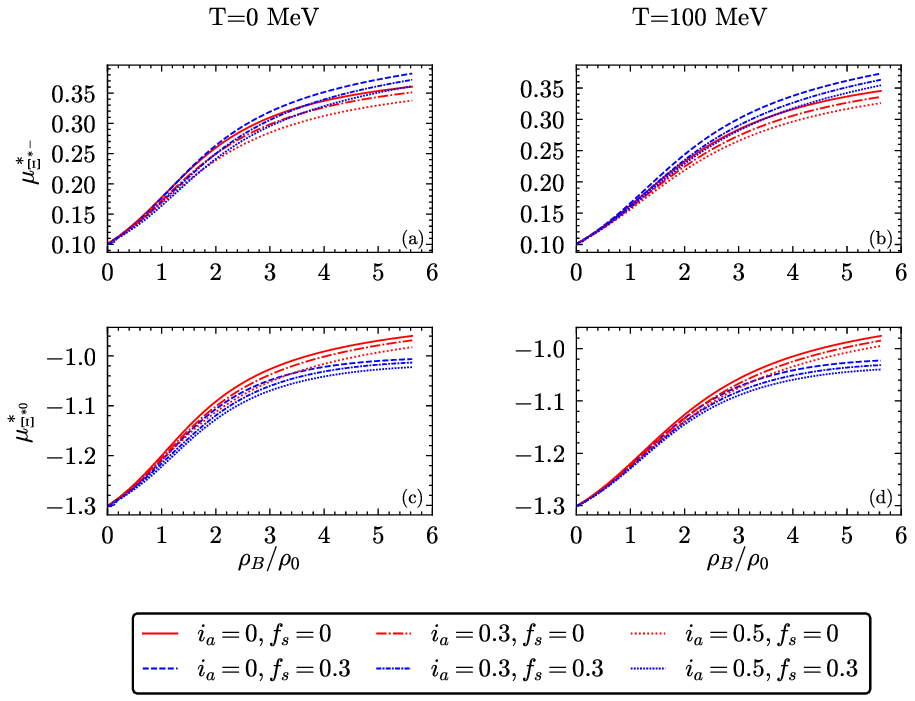}\hfill

	\caption{Variation of effective magnetic moments of $\frac{1}{2}^-$ octet baryon resonances $\Xi^{*-}$ and $\Xi^{*0}$ (in $\mu_N$ units) with baryon density $\rho_B$ (in units of $\rho_0$).}
	\label{fig_total_xi}
\end{figure} 
\begin{figure}[!ht]
	\hspace*{-1.2cm}
	\includegraphics[width=\textwidth, height=16 cm]{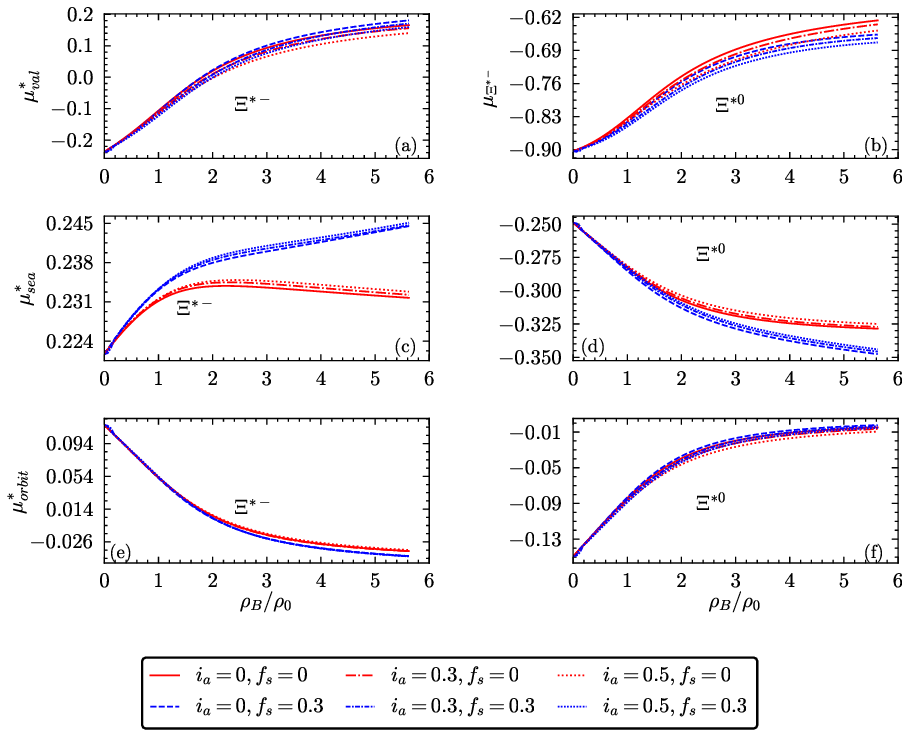}\hfill
		\caption{In-medium trends in explicit contributions to effective magnetic moments of $\frac{1}{2}^-$ octet baryon resonances $\Xi^{*-}$ and $\Xi^{*0}$ (in $\mu_N$ units), from valance quarks ($\mu_{val}^{*}$), sea quarks ($\mu_{sea}^{*}$) and orbital moment of sea quark ($\mu_{orbit}^{*}$) are presented.}
	\label{fig_contri_xi}
\end{figure} 
\begin{figure}[!ht] 
	\hspace*{-1.2cm}
	\includegraphics[width=\textwidth, height=16 cm]{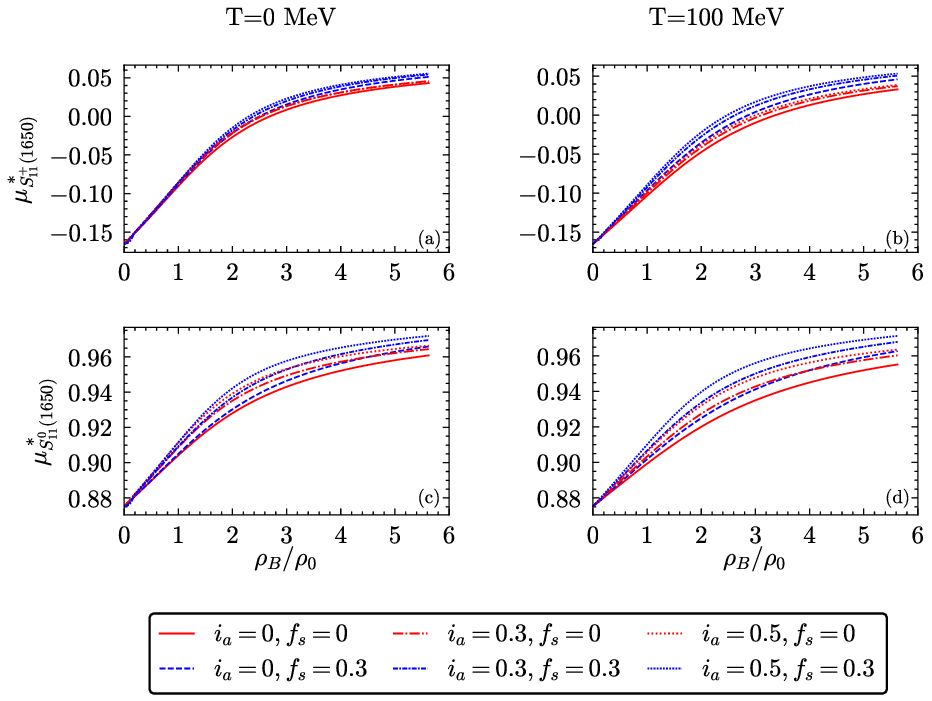}\hfill
	\caption{Variation of effective magnetic moments of $\frac{1}{2}^-$ $N^*$ resonances $S_{11}^{+}(1650)$ and $S_{11}^{0}(1650)$ (in $\mu_N$ units) with baryon density $\rho_B$ (in units of $\rho_0$).}
	\label{fig_total_N2}
\end{figure} 

\begin{figure}[!ht]
	\hspace*{-1.2cm}
	\includegraphics[width=\textwidth, height=16 cm]{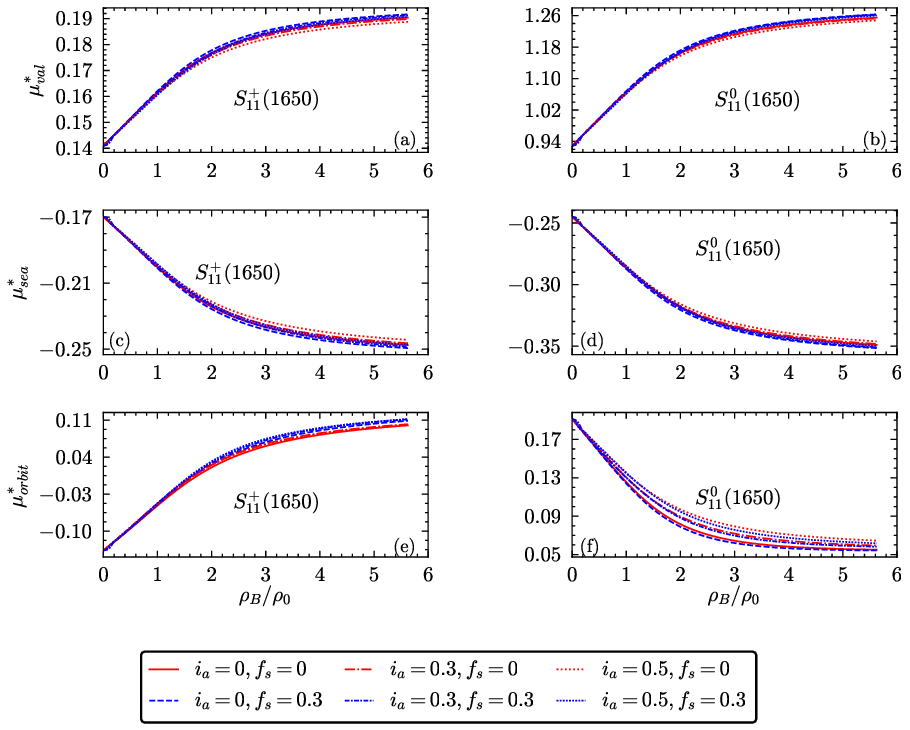}\hfill
	\caption{In-medium trends in explicit contributions to effective magnetic moments of $\frac{1}{2}^-$ $N^*$ resonances $S_{11}^{+}(1650)$ and $S_{11}^{0}(1650)$ (in $\mu_N$ units), from valance quarks ($\mu_{val}^{*}$), sea quarks ($\mu_{sea}^{*}$) and orbital moment of sea quark ($\mu_{orbit}^{*}$) are presented.}
	\label{fig_contri_N2}
\end{figure}

This is where we elaborately examine the results of the vacuum and effective magnetic moments of ${\frac{1}{2}}^-$ octet baryon resonances $p^*, n^*, \Sigma^{*+}, \Sigma^{*-}, \Sigma^{*0}, \Xi^{*-}$, $\Xi^{*0}$ and low-lying $N^*$ resonances ${S^+_{11}}(1650)$ and ${S^0_{11}}(1650)$, observed from \cref{fig_total_pn,fig_contri_pn,fig_total_sigma,fig_contri_sigma,fig_total_xi,fig_contri_xi,fig_total_N2,fig_contri_N2}. For the sake of discussion, we have also included the individual contributions from valence and sea quarks to the total baryonic magnetic moment which allows us to better interpret the results.

In \cref{fig_total_pn}, we have shown the magnetic moment of $p^*$ and $n^*$ resonances with density of the medium at temperatures $T =0$ MeV and $T =100$ MeV. The observed in-medium behavior of $\mu^*_{p^*}$ and $\mu^*_{n^*}$, shows a steeper slope at lower densities of nuclear matter ($\rho_B \leq 3 \rho_0$) and gradually attains saturation at much higher densities. As we trace the plot of these effective baryon magnetic moments (at $T= 0$) while the  symmetric medium density rises to $\rho_B = 3\rho_0$, we observe that $\mu^*_{p^*}$ increases by $0.19 \mu_N$ and $\mu^*_{n^*}$ initially approaches a minima at $-1.382 \mu_N$ (at $\rho_B \approx 1.1\rho_0$) and then climbs up close to its vacuum value at $-1.376 \mu_N$. A conclusive understanding of the observed effective baryonic magnetic moments of $p^*$ and $n^*$ can be provided by analyzing \cref{fig_contri_pn}, where the contributions from valence quarks, sea quarks and orbital angular momentum of the quark sea have been shown individually. For instance, take $n^*$ in symmetric nuclear matter, the decrease in the valence quark contributions dominates over the increasing contributions of sea quarks and the orbital contributions caused due to them, which leads to an overall decreasing behavior of $\mu^*_{n^*}$ below the saturation density of the medium ($\rho_0$). As the medium density moves past this point ($\rho_B \approx \rho_0$), owing to the dominance of $\mu^*_{sea}$ and $\mu^*_{orbit}$ contributions, the effective behavior of $\mu^*_{n^*}$ reverses and it approaches higher values. On a similar note, the density dependency of $\mu^*_{p^*}$ can be comprehended by taking into account the increase in $\mu^*_{val}$ dominating over the decrease in $\mu^*_{sea}$ and $\mu^*_{orbit}$. In symmetric strange matter at $f_s =0.3$, the in-medium behavior of $\mu^*_{p^*}$ and $\mu^*_{n^*}$ is enhanced. From \cref{fig_total_pn} we can infer that, the finite strangeness fraction in the medium leads to a splitting in baryonic magnetic moments of $p^*$ and $n^*$, which is visible beyond the nuclear saturation density, $\rho_0$. All other medium parameters affixed, non-zero isospin asymmetry ($i_a =0, 0.3, 0.5$) causes fine splitting in the $\mu^*_{p^*}$ and $\mu^*_{n^*}$ curves as can be seen from \cref{fig_total_pn}, which has also been identified in their respective $\mu^*_{val}$, $\mu^*_{sea}$, $\mu^*_{orbit}$ contributions by referring to \cref{fig_contri_pn}. As we have included \lq\lq Nambu Goldstone Boson Exchange\rq\rq (NGBE) effect in this study, it is critical to note that the influence of sea quark leads to an opposite sign as compared to the valence quark effect, while on the other hand the orbital angular momentum of sea quarks has the similar contribution as that caused by the valence quarks (see \Cref{table:mub_eta0fs0}). The sign of these contributions can be justified, if we realize that the production of the quark sea is associated with a spin flip process (see \cref{eq: fluctuation}) and hence, the sea quarks has to have an opposite effect. An earlier study highlights the impact of Nambu-Goldstone Boson (NGB) degrees in chiral quark model ($\chi$QM), where the magnetic moments calculated for $p^*$ and $n^*$ in vacuum, by considering NGBE interaction, are 1.4 $\mu_N$ and $-0.9 \mu_N$ for $\chi$QM, which are slightly lower than the corresponding values, 1.7 $\mu_N$ and $-1.1 \mu_N$ as obtained from CQM framework \cite{liu}. In addition, the magnetic moments procured from $\chi$QM by considering Nambu Goldstone Boson Exchange (NGBE) and One Gluon Exchange (OGE), separately, gives similar magnetic moment values and are in close agreement to our vacuum state results \cite{liu}. 
Two other studies in the literature, with quasibound state approach \cite{quasibound} and chiral unitary model \cite{chiral_unitary} have previously calculated for magnetic moments of these two resonances and could be compared from \cref{table:table5}. \Cref{table:table5} facilitates a comparative analysis of the magnetic moments of $p^*$ and $n^*$ resonances obtained from various models.
\Cref{fig_total_sigma} depicts the in-medium modifications of $\mu^*_{\Sigma^{*+}}$, $\mu^*_{\Sigma^{*-}}$ and $\mu^*_{\Sigma^{*0}}$, as a function of $\rho_B / \rho_0$ and the behavior of the individual contributions have been plotted in \cref{fig_contri_sigma}. In nuclear medium, the magnetic moments of $\Sigma^{*+}$ initially increases up to $\rho_B = 2\rho_0$ and then saturates at higher densities of the medium. 
By taking into account the individual contributions from \cref{fig_contri_sigma}, we see that $\mu^*_{val}$ increases with density and it has a dominating effect over the cumulative depletion of $\mu^*_{sea}$ and $\mu^*_{orbit}$ contributions. In case of $\Sigma^{*-}$ (at $i_a=0, f_s=0$), the effective magnetic moment $\mu^*_{\Sigma^{*-}}$ decreases and gradually approaches a saturation at very high densities of $\rho_B \approx 5\rho_0$. Although the in-medium behavior of $\mu^*_{val}$ is accompanied by $\mu^*_{sea}$ and $\mu^*_{orbit}$, which imparts opposite contributions, but the valence quarks have a predominant role in determining the nature of $\mu^*_{\Sigma^{*-}}$. These opposite contributions due to sea quarks and the orbital motion of sea quarks at higher density regime, evidently explains the stagnant in-medium nature of $\mu^*_{\Sigma^{*-}}$ at large medium densities. Further, the density dependent medium modification of $\mu^*_{\Sigma^{*0}}$, in symmetric nuclear matter, is much alike the in-medium nature of $\mu^*_{\Sigma^{*+}}$ except for a slightly decreasing slope at densities above $1.5\rho_0$.
The dominance of $\mu^*_{val}$ and $\mu^*_{orbit}$ contributions over the $\mu^*_{sea}$ contribution, justifies the in-medium curve of $\mu^*_{\Sigma^{*0}}$ in a nuclear medium. 
From a point of comparison if you see \Cref{table:table5}, the magnetic moments of $\mu^*_{\Sigma^{*+}}$ (1.647 $\mu_N$), $\mu^*_{\Sigma^{*-}}$ $(-0.754 \mu_N)$ and $\mu^*_{\Sigma^{*0}}$ (0.446 $\mu_N$) in free space are found to be in agreement with the NCQM results 1.814 $\mu_N$, $-0.689 \mu_N$ and 0.820 $\mu_N$, respectively. As seen from \Cref{table:mub_eta0fs0,table:mub_eta0fs3}, for $f_s =0(0.3)$ at $\rho_B =3\rho_0$, $\mu^*_{\Sigma^{*+}}$ and $\mu^*_{\Sigma^{*-}}$ values attain 1.697(1.723) $\mu_N$ and $-0.770(-0.779) \mu_N$. This suggests that as we introduce finite strangeness fraction to the non-strange medium, there will be an enhancement in the nature of in-medium variations of baryonic magnetic moments of $\Sigma^{*+}$, $\Sigma^{*0}$ and $\Sigma^{*-}$. For finite $f_s$, the magnetic moments of $\Sigma^{*+}$ and $\Sigma^{*0}$ surges while that of $\Sigma^{*-}$ approaches lower values, instead of attaining a saturation at large densities. For an instance, the in-medium values of $\mu^*_{\Sigma^{*+}}$, $\mu^*_{\Sigma^{*-}}$ and $\mu^*_{\Sigma^{*0}}$ (in units of $\mu_N$) at $i_a =0(0.5)$, when referred to \Cref{table:mub_eta0fs0,table:mub_eta5fs0}, are $1.697(1.698), -0.770(-0.763) and 0.463(0.472)$ in a medium density of $3\rho_0$. This and a glance into \cref{fig_total_sigma} makes it apparent that the impact of non-zero $i_a$, a fine splitting in effective baryonic magnetic moments, is more visible for $\mu^*_{\Sigma^{*-}}$ and $\mu^*_{\Sigma^{*0}}$ as compared to $\mu^*_{\Sigma^{*+}}$. Such $i_a$ dependent fine shifts in the baryonic magnetic moment have been previously found in works, in the present formalism, dedicated to octet baryons \cite{fiftythree} and decuplet baryons \cite{fiftyfour}. 

We find it interesting to note that for $\Xi^*$ baryons the density effects on $\mu^*_{\Xi^{*-}}$ and $\mu^*_{\Xi^{*0}}$ values show impressive similarity, owing to the valence quark spin polarizations
and corresponding orbital moments of two $s$ quarks that suppresses the impact of light quarks present in them \cite{sixtytwo}.
 
The baryonic magnetic moment of $\Xi^{*-}$, as seen from \cref{fig_total_xi}, rises steeply with increasing relative hadronic matter density ($\rho_B /\rho_0$). \Cref{fig_contri_xi} clearly illustrates that for $\Xi^{*-}$, the cumulative increasing behavior of $\mu^*_{val}$ and $\mu^*_{sea}$ contributions overshadows the decreasing orbital contributions from sea quarks ($\mu^*_{orbit}$). In a similar manner, a quantitative analysis of \cref{fig_total_xi} and \cref{fig_contri_xi} suggests that the density dependent increase in $\mu^*_{\Xi^{*0}}$ is well justified by the contributions coming from $\mu^*_{val}$ and $\mu^*_{orbit}$, which suppresses the contributions coming from $\mu^*_{sea}$. Much alike other baryons, we have observed a correlation between finite isospin asymmetry of the medium and a distinguishable splitting in the baryonic magnetic moments in case of $\Xi^*$ resonances. Now in order to understand the temperature effects on the in-medium baryonic magnetic moments, we shall lay emphasis on the case of $\mu^*_{\Xi^{*-}}$. A close look at \cref{fig_total_xi}, shows that for symmetric nuclear matter at medium densities of $\rho_0$ and $3\rho_0$, at $T =0(100)$ MeV the $\mu^*_{\Xi^{*-}}$ values obtained, in $\mu_N$ units, are 0.313(0.286) and 0.179(0.163), respectively. This reveals the in-medium nature of $\mu^*_{\Xi^{*-}}$, an increasing curve as a function of medium density, is subject to slight reduction in its value as the temperature rises to $T =100$ MeV. In addition, the vacuum state magnetic moments of $\Xi^{*-}$ resonances are found to be much lower than those procured from NCQM (see \Cref{table:table5}).

From \cref{fig_total_N2}, we observe that the magnetic moments of ${S^+_{11}}(1650)$ and ${S^0_{11}}(1650)$ increases, overall as a function of relative medium density $\rho_B/\rho_0$. As from \cref{fig_contri_N2} it is seen that, the $\mu^*_{sea}$ exercises control over the $\mu^*_{val}$ and $\mu^*_{orbit}$ contributions, which explains the medium modifications of the magnetic moment of ${S^+_{11}}(1650)$ in lower density regime (i.e. up till $\rho_B = 2\rho_0$). However, as the nuclear medium grows denser the $\mu^*_{val}$ and $\mu^*_{orbit}$ contributions demonstrates a dominating influence on the total effective magnetic moment of ${S^+_{11}}(1650)$. From \cref{fig_contri_N2} it is also observed that for the $N^*$ resonance ${S^0_{11}}(1650)$, the value of $\mu_{{S^0_{11}}(1650)}$ receives major contributions, monotonously, due to the dominant valence quark contribution ($\mu^*_{val}$) decreasing with rise in density. For a comparative study based on different models, similar justifications as discussed in the case of $p^*$ and $n^*$ resonances suffices and the results have been tabulated in \Cref{table:table5}. 

A study based on QMC as well as MQMC, for octet baryons, shows that we can expect sizable influence on effective baryonic moments due to temperature above $T = 150$ MeV at lower medium densities \cite{mqmc}. Speaking of correlations, the authors in ref. \cite{seventyfour} have discussed a strong correlation of the baryonic magnetic moment with that of the bag radius of a baryon, as the two quantities have a similar behavior as density dependent functions. Since this is the first work on medium modifications of $\frac{1}{2}^-$ baryon resonances, so referring to previous works on in-medium density effects on magnetic moments of baryons using the current formalism \cite{sixtytwo,fiftyfour} and QMC \cite{tsushima2}, might facilitate the purpose of a comparative study for future work on the baryonic magnetic moments of $\frac{1}{2}^-$ octet resonances.
The effective masses of positive parity octet baryons are also observed to decrease as a function of nuclear density.  As an example, the effective masses of proton $p$ is observed to decrease  by 44 \% as baryon density is decreased from 0 to $3\rho_0$ and this can be compared with the decrease of 25.27\% for negative parity $p^*$ baryon. At high baryon density chiral symmetry is expected to restore
partially and in this limit, properties of positive and negative parity baryons may converge to same values. This kind of picture can be seen more clearly within the present model calculations, if negative parity baryons are considered along with the positive parity baryons in the   solution of system of non-linear equations corresponding to scalar and vector densities of baryons. As we discussed earlier, in the present work properties of negative parity baryons are investigated in strange matter comprising of positive parity nucleons and hyperons.
In our future work, we shall extend our present model calculations to study the thermodynamics of dense matter considering explicitly both positive and negative baryons in
equations of motion of scalar and vector fields through the definitions of scalar and vector densities of baryons.

\section{Summary}
\label{sec:summary}
In a nutshell, this work investigates the medium modified trends induced on the masses and moreover on the magnetic moments of $J_{p}={\frac{1}{2}}^-$ baryon resonances in isospin asymmetric strange hadronic matter. 
The effects caused due to the presence of a medium are introduced through the CQMF model, following which we computed the effective baryon masses by using the masses of constituent quarks. The mass adjusted quark magnetic moments are then calculated, to include relativistic quark confinement, by using the baryon masses as inputs. In hadronic medium the in-medium masses of lights quarks, $u$ \& $d$, vary significantly in comparison to the observed changes in medium modified masses of $s$ quark. In the $\chi$CQM formalism, we then obtained the baryonic magnetic moments by the summation over the contributed values coming from valence quarks, sea quarks and orbital angular momentum due of sea quarks. As derived from the results is that, the medium modifications are more prominent for the baryons comprised of light $u$ or $d$ quarks, than for hyperons. For strange matter, the in-medium masses of constituent quark and baryons as well as the effective baryonic magnetic moments suffer appreciable enhancement in their in-medium behaviour. By carrying out this study at finite temperatures and finite $i_a$ values of the hadronic medium, we have noted a fine shift in the in-medium baryonic magnetic moments, although the temperature effects have been negligible below $T = 150$ MeV.

The results from this study could be a point of reference to experiments which study hadron spectroscopy by probing photoproductions. In Jefferson Lab the GlueX experiment has measured the width of the N(1535) state to be narrower (less than 100 MeV) making it undetetctable, whereas its cascade program has aligned its goals to explore the negative parity excited $\Xi$ baryons\cite{baryon_spectroscopy}. In the limelight of excited cascade resonances, a report from the Belle Collaboration sheds light on the detection of two states in low mass regime of which one ought to be $\Xi$(1690) \cite{Belle}. Such experimental studies have research interests directed towards a detailed understanding of QCD for low-lying hadronic structures. Hence, the density dependent study of negative parity first excited states of octet baryon resonances is a crucial point of experimental interest at exploring the predictions of constituent quark models. Additionally, this work has scope for further extrapolation in the context of higher density astrophysical objects like neutron stars, to understand the impact of a dense medium on the baryonic magnetic moments \cite{medium_effect_neutron}. 

\begin{table}[ht]
	\centering
	\renewcommand{\arraystretch}{1.5}	
	\[
\begin{array}{|c|c|c|c|c|c|c|}
	\hline
	k_0 & k_1 & k_2 & k_3 & k_4 \\ \hline
	4.94 & 2.12 & -10.16 & -5.38 & -0.06 \\ \hline
	\sigma_0 \, (\text{MeV}) & \quad \zeta_0 \, (\text{MeV}) & \quad \chi_0 \, (\text{MeV}) & \quad \xi & \quad \rho_0 \, (\text{fm}^{-3}) \\ \hline
	-92.8 & -96.5 & 254.6 & \frac{6}{33} & 0.16 \\ \hline
	g_\sigma^u = g_\sigma^d & \quad g_\sigma^s = g_\zeta^u = g_\zeta^d & \quad g_\delta^u & g_\zeta^s = g_s & g_4 \\ \hline
	2.72 & 0 & 2.72 & 3.847 & 37.4 \\ \hline
	g_\rho^p = g_\rho^u & g_\omega^N = 3 g_\omega^u & \quad g_\rho^p & \quad m_\pi \, (\text{MeV}) & \quad m_K \, (\text{MeV}) \\ \hline
	2.72 & 9.69 & 8.886 & 139 & 494 \\ \hline
\end{array}
\]

\caption{In above table various parameters used in the current manuscript are summarized.}
\label{tab:array-table}
\end{table}

	\begin{sidewaystable}[h!]
		\resizebox{\textwidth}{!}{%
			\begin{tabular}{|p{1.62cm}|p{2.4cm}|c|c|c|c|c|c|c|c|}
				\hline
				\multirow{3}{*}{Baryon} & \multirow{3}{*}{Vacuum Mass} & \multicolumn{4}{c|}{$\rho_{B}=\rho_{0}$} & \multicolumn{4}{c|}{$\rho_{B}=3\rho_{0}$} \\ \cline{3-10}
				& & \multicolumn{2}{c|}{$f_s=0$} & \multicolumn{2}{c|}{$f_s=0.3$} & \multicolumn{2}{c|}{$f_s=0$} & \multicolumn{2}{c|}{$f_s=0.3$} \\ \cline{3-10}
				& (MeV) & $i_a=0$ & $i_a=0.3$ &  $i_a=0$ &  $i_a=0.3$ &  $i_a=0$ &  $i_a=0.3$ &  $i_a=0$ &  $i_a=0.3$ \\
				\hline 
				$M_{ S^{+}_{11}(1650) }^*$ &  1650.0 &  $1439.01$ &  $1442.02$ &  $1435.09$ &  $1440.78$ &  $1263.40$ &  $1268.27$ &  $1254.90$ &  $1260.31$ \\
				\hline $M_{ S^{0}_{11}(1650) }^*$ &  1650.0 &  $1439.01$ &  $1439.98$ &  $1435.09 $&  $1438.69 $&  $1263.40 $&  $1265.55 $&  $1254.90 $&  $1257.48$ \\
				\hline $M_{ p^{*} }^*$ &  1535.0 &  $1323.31 $&  $1326.33 $&  $1319.37 $&  $1325.08 $&  $1147.02 $&  $1151.91 $&  $1138.50 $&  $1143.92$ \\
				\hline $M_{ n^{*} }^*$ &  1535.0 &  $1323.31 $&  $1324.28 $&  $1319.37 $&  $1322.98 $&  $1147.02 $&  $1149.18 $&  $1138.50 $&  $1141.08$ \\
				\hline $M_{ \Sigma^{*+} }^*$ &  1620.0 &  $1442.84 $&  $1446.59 $&  $1434.69 $&  $1440.74 $&  $1300.88 $&  $1306.54 $&  $1276.30 $&  $1283.18$ \\
				\hline $M_{ \Sigma^{*-} }^*$ &  1620.0 &  $1442.84 $&  $1442.48 $&  $1434.69 $&  $1436.54 $&  $1300.88 $&  $1301.06 $&  $1276.30 $&  $1277.49$ \\
				\hline $M_{ \Sigma^{*0} }^*$ &  1620.0 &  $1442.84 $&  $1444.53 $&  $1434.69 $&  $1438.64 $&  $1300.88 $&  $1303.80 $&  $1276.30 $&  $1280.34$ \\
				\hline $M_{ \Xi^{*-} }^*$ &  1690.0 &  $1547.34 $&  $1547.70 $&  $1534.99 $&  $1537.17 $&  $1439.75 $&  $1440.69 $&  $1399.14 $&  $1401.79$ \\
				\hline $M_{ \Xi^{*0} }^*$ &  1690.0 &  $1547.34 $&  $1549.75 $&  $1534.99 $&  $1539.27 $&  $1439.75 $&  $1443.43 $&  $1399.14 $&  $1404.64$ \\
				\hline
			\end{tabular}
		}
		\caption{The  effective masses of spin$-\frac{1}{2}^-$ octet baryon resonances are tabulated at $\rho_B = \rho_0$ and $3\rho_0$ for different values of isospin asymmetry ($i_a$) and strangeness fraction ($f_s$) at $T = 0$ MeV.}
		\label{tab:masses_T0}
	\end{sidewaystable}

%
\begin{sidewaystable}[h!]
\begin{tabular}{|p{1.62cm}|p{1.4cm}|c|c|c|c|c|c|c|c|}
\hline
	\multirow{3}{*}{Baryon} & \multirow{3}{*}{Vacuum Mass   
					} & \multicolumn{4}{c|}{$\rho_{B}=\rho_{0}$} & \multicolumn{4}{c|}{$\rho_{B}=3\rho_{0}$} \\ \cline{3-10}
			&  & \multicolumn{2}{c|}{$f_s=0$} & \multicolumn{2}{c|}{$f_s=0.3$} & \multicolumn{2}{c|}{$f_s=0$} & \multicolumn{2}{c|}{$f_s=0.3$} \\ \cline{3-10}
			&  (MeV)  & $i_a=0$ & $i_a=0.3$ &  $i_a=0$ &  $i_a=0.3$ &  $i_a=0$ &  $i_a=0.3$ &  $i_a=0$ &  $i_a=0.3$ \\  
		\hline $M_{ S^{+}_{11}(1650) }^*$ &  1650.0 &  $1469.66$ &  $1471.49$ &  $1454.68$ &  $1455.95$ &  $1289.10$ &  $1292.78$ &  $1270.06$ &  $1272.66$ \\
		\hline $M_{ S^{0}_{11}(1650) }^*$ &  1650.0 &  $1469.66$ &  $1469.70$ &  $1454.68 $&  $1453.87 $&  $1289.10 $&  $1289.73 $&  $1270.06 $&  $1269.38$ \\
		\hline $M_{ p^{*} }^*$ &  1535.0 & $1354.07 $&  $1355.90 $&  $1339.04 $&  $1340.30 $&  $1172.82 $&  $1176.51 $&  $1153.71 $&  $1156.32$ \\
		\hline $M_{ n^{*} }^*$ &  1535.0 &  $1354.07 $&  $1354.10 $&  $1339.04 $&  $1338.22 $&  $1172.82 $&  $1173.45 $&  $1153.71 $&  $1153.02$ \\
		\hline $M_{ \Sigma^{*+} }^*$ &  1620.0 &  $1468.45 $&  $1471.06 $&  $1450.57 $&  $1452.89 $&  $1322.58 $&  $1327.39 $&  $1287.69 $&  $1291.80$ \\
		\hline $M_{ \Sigma^{*-} }^*$ &  1620.0 &  $1468.45 $&  $1467.46 $&  $1450.57 $&  $1448.72 $&  $1322.58 $&  $1321.26 $&  $1287.69 $&  $1285.20$ \\
		\hline $M_{ \Sigma^{*0} }^*$ &  1620.0 &  $1468.45 $&  $1469.26 $&  $1450.57 $&  $1450.81 $&  $1322.58 $&  $1324.33 $&  $1287.69 $&  $1288.50$ \\
		\hline $M_{ \Xi^{*-} }^*$ &  1690.0 &  $1567.79 $&  $1567.58 $&  $1547.08 $&  $1546.28 $&  $1457.33 $&  $1457.12 $&  $1406.70 $&  $1405.71$ \\
		\hline $M_{ \Xi^{*0} }^*$ &  1690.0 &  $1567.79 $&  $1569.38 $&  $1547.08 $&  $1548.36 $&  $1457.33 $&  $1460.19 $&  $1406.70 $&  $1409.02$ \\

		\hline
\end{tabular}
		\caption{The shift in effective masses of spin$-\frac{1}{2}^-$ octet baryons with respect to vacuum masses are tabulated at $\rho_B = \rho_0$ and $3\rho_0$ for isospin asymmetry $i_a = 0,\ 0.3$ and strangeness fractions $f_s = 0,\ 0.3$ at temperatures $T = 100$ MeV.}
		\label{tab:masses_T100}
		\end{sidewaystable}
		
		\begin{sidewaystable}
			\resizebox{\textwidth}{!}{%
				\begin{tabular}{|c|c|c|c|c|c|c|c|c|c|c|c|c|}
					\hline 
					\multirow{2}{*}{Baryons} & \multicolumn{4}{|c|}{$\rho_B=0$} & \multicolumn{4}{|c|}{$\rho_B=\rho_0$} & \multicolumn{4}{|c|}{$\rho_B=3 \rho_0$} \\
					\cline{2-13} 
					& $\mu^*_{\text{val}}$ & $\mu^*_{\text{sea}}$ & $\mu^*_{\text{orbit}}$ & $\mu_B^*$ & $\mu^*_{\text{val}}$ & $\mu^*_{\text{sea}}$ & $\mu^*_{\text{orbit}}$ & $\mu_B^*$ & $\mu^*_{\text{val}}$ & $\mu^*_{\text{sea}}$ & $\mu^*_{\text{orbit}}$ & $\mu_B^*$ \\
					\hline
					${S^{+}_{11}(1650)}\left(\mu_N \right)$ &
					0.140 & -0.169 & -0.136 & -0.165 &
					0.161 & -0.200 & -0.051 & -0.090 &
					0.184 & -0.236 & 0.061 & 0.008 \\
					\hline
					${S^{0}_{11}(1650)}\left(\mu_N \right)$ &
					0.929 & -0.244 & 0.191 & 0.875 &
					1.065 & -0.286 & 0.125 & 0.904 &
					1.213 & -0.334 & 0.064 & 0.943 \\
					\hline
					${p}^*\left(\mu_N \right)$ &
					1.859 & -0.318 & 0.349 & 1.889 &
					2.156 & -0.368 & 0.194 & 1.982 &
					2.487 & -0.423 & 0.014 & 2.078 \\
					\hline
					${n}^*\left(\mu_N \right)$ &
					-1.262 & 0.226 & -0.339 & -1.375 &
					-1.464 & 0.263 & -0.181 & -1.382 &
					-1.689 & 0.305 & 0.007 & -1.376 \\
					\hline
					${\Sigma}^{*+}\left(\mu_N \right)$ &
					1.761 & -0.343 & 0.229 & 1.647 &
					1.935 & -0.382 & 0.132 & 1.684 &
					2.086 & -0.419 & 0.030 & 1.697 \\
					\hline
					${\Sigma}^{*-}\left(\mu_N \right)$ &
					-0.694 & 0.133 & -0.193 & -0.754 &
					-0.821 & 0.153 & -0.091 & -0.760 &
					-0.971 & 0.174 & 0.026 & -0.770 \\
					\hline
					${\Sigma}^{*0}\left(\mu_N \right)$ &
					0.533 & -0.105 & 0.017 & 0.446 &
					0.556 & -0.114 & 0.020 & 0.461 &
					0.557 & -0.122 & 0.028 & 0.463 \\
					\hline
					${\Xi}^{*-}\left(\mu_N\right)$ &
					-0.237 & 0.221 & 0.116 & 0.100 &
					-0.106 & 0.231 & 0.052 & 0.177 &
					0.093 & 0.233 & -0.017 & 0.309 \\
					\hline
					${\Xi}^{*0}\left(\mu_N\right)$ &
					-0.903 & -0.248 & -0.149 & -1.301 &
					-0.833 & -0.282 & -0.084 & -1.200 &
					-0.687 & -0.318 & -0.019 & -1.026 \\
					\hline
			\end{tabular}}
			\caption{Computed values of effective magnetic moments of $\frac{1}{2}^-$ baryon resonances at different medium densities ($\rho_B / \rho_0 =0, 1, 3$) are tabulated in the presence of symmetric nuclear matter ($i_a =0$ \& $f_s =0$) at $T=0$ MeV.}
			\label{table:mub_eta0fs0}
		\end{sidewaystable}

		\begin{sidewaystable}
			\resizebox{\textwidth}{!}{%
				\begin{tabular}{|c|c|c|c|c|c|c|c|c|c|c|c|c|}
					\hline 
					\multirow{2}{*}{Baryons} & \multicolumn{4}{|c|}{$\rho_B=0$} & \multicolumn{4}{|c|}{$\rho_B=\rho_0$} & \multicolumn{4}{|c|}{$\rho_B=3 \rho_0$} \\
					\cline{2-13} 
					& $\mu^*_{\text{val}}$ & $\mu^*_{\text{sea}}$ & $\mu^*_{\text{orbit}}$ & $\mu_B^*$ & $\mu^*_{\text{val}}$ & $\mu^*_{\text{sea}}$ & $\mu^*_{\text{orbit}}$ & $\mu_B^*$ & $\mu^*_{\text{val}}$ & $\mu^*_{\text{sea}}$ & $\mu^*_{\text{orbit}}$ & $\mu_B^*$ \\
					\hline
					${S^{+}_{11}(1650)}\left(\mu_N \right)$ &
					0.140 & -0.169 & -0.136 & -0.165 &
					0.160 & -0.199 & -0.047 & -0.085 &
					0.182 & -0.233 & 0.064 & 0.013 \\
					\hline
					${S^{0}_{11}(1650)}\left(\mu_N \right)$ &
					0.929 & -0.244 & 0.191 & 0.875 &
					1.062 & -0.285 & 0.134 & 0.911 &
					1.205 & -0.331 & 0.079 & 0.953 \\
					\hline
					${p}^*\left(\mu_N \right)$ &
					1.859 & -0.318 & 0.349 & 1.889 &
					2.151 & -0.367 & 0.196 & 1.980 &
					2.477 & -0.421 & 0.019 & 2.074 \\
					\hline
					${n}^*\left(\mu_N \right)$ &
					-1.262 & 0.226 & -0.339 & -1.375 &
					-1.463 & 0.263 & -0.183 & -1.382 &
					-1.686 & 0.305 &  0.004 & -1.376 \\
					\hline
					${\Sigma}^{*+}\left(\mu_N \right)$ &
					1.761 & -0.343 & 0.229 & 1.647 &
					1.927 & -0.380 & 0.137 & 1.684 &
					2.072 & -0.416 & 0.042 & 1.698 \\
					\hline
					${\Sigma}^{*-}\left(\mu_N \right)$ &
					-0.694 & 0.133 & -0.193 & -0.754 &
					-0.818 & 0.152 & -0.092 & -0.757 &
					-0.960 & 0.173 & 0.024 & -0.763 \\
					\hline
					${\Sigma}^{*0}\left(\mu_N \right)$ &
					0.533 & -0.105 & 0.017 & 0.446 &
					0.557 & -0.114 & 0.023 & 0.466 &
					0.561 & -0.122 & 0.033 & 0.472 \\
					\hline
					${\Xi}^{*-}\left(\mu_N\right)$ &
					-0.237 & 0.221 & 0.116 & 0.100 &
					-0.116 & 0.231 & 0.052 & 0.168 &
					0.065 & 0.234 & -0.015 & 0.284 \\
					\hline
					${\Xi}^{*0}\left(\mu_N\right)$ &
					-0.903 & -0.248 & -0.149 & -1.301 &
					-0.840 & -0.281 & -0.087 & -1.209 &
					-0.710 & -0.314 & -0.026 & -1.051 \\
					\hline
			\end{tabular}}
			\caption{Computed values of effective magnetic moments of $\frac{1}{2}^-$ baryon resonances at different medium densities ($\rho_B / \rho_0 =0, 1, 3$) are tabulated in the presence of asymmetric nuclear matter ($i_a =0.5$ \& $f_s =0$) at  $T=0$ MeV.}
			\label{table:mub_eta5fs0}
		\end{sidewaystable}
		
		\begin{sidewaystable}
			\resizebox{\textwidth}{!}{
				\begin{tabular}{|c|c|c|c|c|c|c|c|c|c|c|c|c|}
					\hline 
					\multirow{2}{*}{Baryons} & \multicolumn{4}{|c|}{$\rho_B=0$} & \multicolumn{4}{|c|}{$\rho_B=\rho_0$} & \multicolumn{4}{|c|}{$\rho_B=3 \rho_0$} \\
					\cline{2-13} 
					& $\mu^*_{\text{val}}$ & $\mu^*_{\text{sea}}$ & $\mu^*_{\text{orbit}}$ & $\mu_B^*$ & $\mu^*_{\text{val}}$ & $\mu^*_{\text{sea}}$ & $\mu^*_{\text{orbit}}$ & $\mu_B^*$ & $\mu^*_{\text{val}}$ & $\mu^*_{\text{sea}}$ & $\mu^*_{\text{orbit}}$ & $\mu_B^*$ \\
					\hline
					${S^{+}_{11}(1650)}\left(\mu_N \right)$ &
					0.140 & -0.169 & -0.136 & -0.165 &
					0.162 & -0.200 & -0.049 & -0.088 &
					0.185 & -0.238 & 0.068 & 0.015 \\
					\hline
					${S^{0}_{11}(1650)}\left(\mu_N \right)$ &
					0.929 & -0.244 & 0.191 & 0.875 &
					1.068 & -0.287 & 0.124 & 0.904 &
					1.221 & -0.337 & 0.062 & 0.946 \\
					\hline
					${p}^*\left(\mu_N \right)$ &
					1.859 & -0.318 & 0.349 & 1.889 &
					2.162 & -0.369 & 0.191 & 1.984 &
					2.506 & -0.426 & 0.004 & 2.084 \\
					\hline
					${n}^*\left(\mu_N \right)$ &
					-1.262 & 0.226 & -0.339 & -1.375 &
					-1.468 & 0.264 & -0.178 & -1.382 &
					-1.701 & 0.308 & 0.018 & -1.375 \\
					\hline
					${\Sigma}^{*+}\left(\mu_N \right)$ &
					1.761 & -0.343 & 0.229 & 1.647 &
					1.946 & -0.384 & 0.130 & 1.692 &
					2.125 & -0.427 & 0.026 & 1.723 \\
					\hline
					${\Sigma}^{*-}\left(\mu_N \right)$ &
					-0.694 & 0.133 & -0.193 & -0.754 &
					-0.826 & 0.153 & -0.089 & -0.762 &
					-0.991 & 0.177 & 0.034 & -0.779 \\
					\hline
					${\Sigma}^{*0}\left(\mu_N \right)$ &
					0.533 & -0.105 & 0.017 & 0.446 &
					0.560 & -0.115 & 0.020 & 0.464 &
					0.567 & -0.125 & 0.030 & 0.472 \\
					\hline
					${\Xi}^{*-}\left(\mu_N\right)$ &
					-0.237 & 0.221 & 0.116 & 0.100 &
					-0.108 & 0.233 & 0.051 & 0.176 &
					0.100 & 0.239 & -0.021 & 0.319 \\
					\hline
					${\Xi}^{*0}\left(\mu_N\right)$ &
					-0.903 & -0.248 & -0.149 & -1.301 &
					-0.841 & -0.285 & -0.083 & -1.209 &
					-0.703 & -0.328 & -0.016 & -1.048 \\
					\hline
			\end{tabular}}
			\caption{Computed values of effective magnetic moments of $\frac{1}{2}^-$ baryon resonances at different medium densities ($\rho_B / \rho_0 =0, 1, 3$) are tabulated in the presence of symmetric hyperonic matter ($i_a =0$ \& $f_s =0.3$) at $T=0$ MeV.}
			\label{table:mub_eta0fs3}
		\end{sidewaystable}

		\begin{sidewaystable}
			\resizebox{\textwidth}{!}{%
				\begin{tabular}{|c|c|c|c|c|c|c|c|c|c|c|c|c|}
					\hline 
					\multirow{2}{*}{Baryons} & \multicolumn{4}{|c|}{$\rho_B=0$} & \multicolumn{4}{|c|}{$\rho_B=\rho_0$} & \multicolumn{4}{|c|}{$\rho_B=3 \rho_0$} \\
					\cline{2-13} 
					& $\mu^*_{\text{val}}$ & $\mu^*_{\text{sea}}$ & $\mu^*_{\text{orbit}}$ & $\mu_B^*$ & $\mu^*_{\text{val}}$ & $\mu^*_{\text{sea}}$ & $\mu^*_{\text{orbit}}$ & $\mu_B^*$ & $\mu^*_{\text{val}}$ & $\mu^*_{\text{sea}}$ & $\mu^*_{\text{orbit}}$ & $\mu_B^*$ \\

					\hline
					${S^{+}_{11}(1650)}\left(\mu_N \right)$ &
					0.140 & -0.169 & -0.136 & -0.165 &
					0.160 & -0.198 & -0.047 & -0.085 &
					0.183 & -0.235 & 0.074 & 0.022 \\
					\hline
					${S^{0}_{11}(1650)}\left(\mu_N \right)$ &
					0.929 & -0.244 & 0.191 & 0.875 &
					1.061 & -0.284 & 0.135 & 0.911 &
					1.216 & -0.334 & 0.075 & 0.957 \\
					\hline
					${p}^*\left(\mu_N \right)$ &
					1.859 & -0.318 & 0.349 & 1.889 &
					2.142 & -0.365 & 0.201 & 1.977 &
					2.485 & -0.423 & 0.013 & 2.075 \\
					\hline
					${n}^*\left(\mu_N \right)$ &
					-1.262 & 0.226 & -0.339 & -1.375 &
					-1.458 & 0.262 & -0.187 & -1.382 &
					-1.694 & 0.306 & 0.012 & -1.375 \\
					\hline
					${\Sigma}^{*+}\left(\mu_N \right)$ &
					1.761 & -0.343 & 0.229 & 1.647 &
					1.933 & -0.381 & 0.138 & 1.690 &
					2.112 & -0.424 & 0.035 & 1.723 \\
					\hline
					${\Sigma}^{*-}\left(\mu_N \right)$ &
					-0.694 & 0.133 & -0.193 & -0.754 &
					-0.818 & 0.152 & -0.093 & -0.758 &
					-0.982 & 0.176 & 0.034 & -0.771 \\
					\hline
					${\Sigma}^{*0}\left(\mu_N \right)$ &
					0.533 & -0.105 & 0.017 & 0.446 &
					0.560 & -0.115 & 0.023 & 0.468 &
					0.570 & -0.124 & 0.034 & 0.480 \\
					\hline
					${\Xi}^{*-}\left(\mu_N\right)$ &
					-0.237 & 0.221 & 0.116 & 0.100 &
					-0.122 & 0.233 & 0.053 & 0.163 &
					0.076 & 0.240 & -0.021 & 0.295 \\
					\hline
					${\Xi}^{*0}\left(\mu_N\right)$ &
					-0.903 & -0.248 & -0.149 & -1.301 &
					-0.851 & -0.282 & -0.088 & -1.222 &
					-0.722 & -0.324 & -0.022 & -1.069 \\
					\hline
			\end{tabular}}
			\caption{Computed values of in-medium magnetic moments of $\frac{1}{2}^-$ baryon resonances at different medium densities ($\rho_B / \rho_0 =0, 1, 3$) are tabulated in the presence of asymmetric hyperonic matter ($i_a =0.5$ \& $f_s =0.3$) at $T=0$ MeV.}
			\label{table:mub_eta5fs3}
		\end{sidewaystable}

			\begin{sidewaystable}
				\centering
				\resizebox{\textwidth}{!}{%
					\begin{tabular}{|c|c|c|c|c|c|c|c|c|c|c|}
						\hline 
						\multirow{2}{*}{Baryons} & \multicolumn{4}{|c|}{$\chi$CQM} & \multicolumn{2}{|c|}{$\chi$QM\tiny{\cite{liu}}} & \multirow{1}{*}{CQM\tiny{\cite{liu}}} & \multirow{1}{*}{NCQM\tiny{\cite{thirtyone}}} & \multirow{1}{*}{Quasibound\tiny{\cite{quasibound}}} & \multirow{1}{*}{CU\tiny{\cite{chiral_unitary}}}  \\
						\cline{2-11} 
						& $\mu_{\text{val}}$ & $\mu_{\text{sea}}$ & $\mu_{\text{orbit}}$ & $\mu_B$ & $\mu_B$ \tiny(NGBE) & $\mu_B$ \tiny(OGE) & $\mu_B$ \tiny(NGBE) & $\mu_B$ & $\mu_B$ & $\mu_B$ \\
						\hline
						${S^{+}_{11}(1650)}\left(\mu_N \right)$ &
						0.140 & -0.169 & -0.136 & -0.165 & 
						0.1 & 0.0 & 0.3 & 0.106 & - & - \\
						\hline
						${S^{0}_{11}(1650)}\left(\mu_N \right)$ &
						0.929 & -0.244 & 0.191 & 0.875 &
						0.6 & 0.7 & 0.8 & 0.951 & - & - \\
						\hline
						${p}^*\left(\mu_N \right)$ &
						1.859 & -0.318 & 0.349 & 1.889 &
						1.4 & 1.6 & 1.7 & 1.894 & 1.86 & 1.13 \\
						\hline
						${n}^*\left(\mu_N \right)$ &
						-1.262 & 0.226 & -0.339 & -1.375 &
						-0.9 & -1.0 & -1.1 & -1.284 & -0.56 & -0.248\\
						\hline
						${\Sigma}^{*+}\left(\mu_N \right)$ &
						1.761 & -0.343 & 0.229 & 1.647 &
						- & - & - & 1.814 & - & -  \\
						\hline
						${\Sigma}^{*-}\left(\mu_N \right)$ &
						-0.694 & 0.133 & -0.193 & -0.754 &
						- & - & - & -0.689 & - & -  \\
						\hline
						${\Sigma}^{*0}\left(\mu_N \right)$ &
						0.533 & -0.105 & 0.017 & 0.446 &
						- & - & - & 0.820 & - & - \\
						\hline
						${\Xi}^{*-}\left(\mu_N\right)$ &
						-0.237 & 0.221 & 0.116 & 0.100 &
						- & - & - & -0.315 & - & - \\
						\hline
						${\Xi}^{*0}\left(\mu_N\right)$ &
						-0.903 & -0.248 & -0.149 & -1.301 &
						- & - & - & -0.990 & - & - \\
						\hline
					\end{tabular}
				}
				\caption{The above table allows a comparative study of magnetic moments of $\frac{1}{2}^-$ baryon resonances, computed using different models in the free space ($\rho_B = 0$). 
				}
				\label{table:table5}
			\end{sidewaystable}

\end{document}